\documentclass[conference]{IEEEtran}
\IEEEoverridecommandlockouts
\usepackage{cite}
\usepackage{amsmath,amssymb,amsfonts}
\usepackage{algorithmic}
\usepackage{algorithm2e}
\usepackage{graphicx}
\usepackage{textcomp}
\usepackage{xcolor}
\def\BibTeX{{\rm B\kern-.05em{\sc i\kern-.025em b}\kern-.08em
    T\kern-.1667em\lower.7ex\hbox{E}\kern-.125emX}}

\usepackage{pgfplots}
\usepackage{float}
\pgfplotsset{compat=newest}
\usepackage{subcaption}
\usepackage{url}
\usepackage{multirow}
\usepackage{booktabs}

\newcommand{\ex}{{\mathbb{E}}}
\newcommand{\var}{{\mathbb{V}ar}}

\newcommand{\pr}{{\mathbb{P}}}
\newcommand{\cD}{{\mathcal{D}}}    
    
\begin{document}

\title{Solving sparse linear systems with approximate inverse 
preconditioners on analog devices}

\author{\IEEEauthorblockN{Vasileios Kalantzis\IEEEauthorrefmark{1},
Anshul Gupta\IEEEauthorrefmark{2}, 
Lior Horesh\IEEEauthorrefmark{2},
Tomasz Nowicki\IEEEauthorrefmark{2}, 
Mark S.\ Squillante\IEEEauthorrefmark{2}, and
Chai Wah Wu\IEEEauthorrefmark{2}}
\IEEEauthorblockA{IBM Research, Thomas J. Watson Research Center, Yorktown Heights, NY, USA \\
Email: \IEEEauthorrefmark{1}vkal@ibm.com,
\IEEEauthorrefmark{2}\{anshul,lhoresh,tnowicki,mss,cwwu\}@us.ibm.com
}}

\maketitle

\begin{abstract}
Sparse linear system solvers are computationally expensive kernels that lie
at the heart of numerous applications.
This paper proposes 
a preconditioning framework that combines approximate inverses with stationary iterations to substantially reduce the time and energy requirements of this task by utilizing a hybrid architecture that combines conventional digital microprocessors with analog crossbar array accelerators.
Our analysis
and experiments with a simulator for analog hardware demonstrate that an order of magnitude speedup is readily attainable without much impact on convergence, despite the noise in analog computations.

\end{abstract}


\section{Introduction}
\label{sec-intro}

The iterative solution of sparse linear systems~\cite{saad2003iterative} is a fundamental task across many applications in science, engineering, and optimization. For decades, progress in speeding up preconditioned iterative solvers 
was primarily driven by constructing more effective preconditioning algorithms and via steady microprocessor performance gains.
The last decade saw the advent of fast power-efficient low-precision hardware, such as 
Graphics Processing Units (GPUs), which gave rise to new opportunities to accelerate 
sparse iterative solvers~\cite{bertaccini2016sparse,abdelfattah2020survey,oo2020accelerating,baboulin2009accelerating,haidar2018harnessing,anzt2017preconditioned,haidar2020mixed}.
While
these advances
have led to remarkable gains in the performance of sparse iterative solvers, 
conventional CMOS-based digital microprocessors 
possess inherent scaling and power dissipation limitations. 
It is therefore imperative that we explore alternative hardware architectures 
to address current and future challenges of sparse iterative 
solvers. One such alternative paradigm is based on analog devices configured as crossbar arrays of non-volatile memory.
Such arrays can achieve high degrees of parallelism with low energy 
consumption by mapping matrices onto arrays of memristive elements capable of 
storing information and executing simple operations such as a multiply-and-add. 
In particular, these devices can perform Matrix-Vector Multiplication (MVM) in near constant time independent 
of the number of nonzero entries in the operand matrices~\cite{hu2016dot,xia2016technological}. Indeed, considerable computational speed-ups have been achieved with analog crossbar arrays~\cite{Sebastian-NN,Haensch2019,RAPA2019,Ambrogio2018,Fumorola2016,Gokmen2016,Burr2016,Bojnordi2016,Shafice2016}, mostly for machine learning applications that can tolerate the noise and lower precision of these devices.

Analog designs with enhanced precision have been proposed~\cite{Feinberg2021,Feinberg2018} to meet the accuracy demands of scientific applications, but the improvement in precision comes at the cost of increased hardware complexity and energy consumption.
Memristive hardware with precision enhancement has been 
exploited in the context of Jacobi iterations to solve linear systems stemming from
partial differential equations (PDEs)~\cite{zidan2018general}, and as an autonomous linear system solver~\cite{sun2019solving} for small dense systems. The latter approach was used to apply overlapping block-Jacobi preconditioners in the
Generalized Minimal RESidual (GMRES)
iterative solver 
\cite{Feinberg2021} with bit-slicing for increasing precision. Inner-outer iterations~\cite{richter2015memristive,LeGallo2018,zhu2021fixed} have been considered for dense systems, where analog arrays were used for the inner solver while 
iterative refinement was used as an outer solver in the digital space. 

The main contribution of this paper is a flexible preconditioning framework for inexpensively solving a large class of sparse linear systems by effectively utilizing simple analog crossbar arrays without precision-enhancing extensions.
Our approach combines flexible iterative solvers and approximate inverse preconditioners on a hybrid architecture consisting of a conventional digital processor and an analog crossbar array accelerator.
We analytically and experimentally show that this combination can solve certain sparse linear systems at a fraction of the cost of solving them on digital processors. While the relative advantage of analog hardware grows with the size of the linear systems, we demonstrate the potential for an order of magnitude speedup even for systems with just a few hundred equations.

While our method can be used to provide a fast stand-alone sparse solver, it also has exciting applications in the context of the upcoming exascale platforms~\cite{Exascale2020}. Extreme-scale solvers are likely to be highly composable~\cite{Composable2012} and hierarchical~\cite{Parco2014} with multiple levels of nested algorithmic components. A solver like the one proposed in this paper, running on analog accelerator-equipped nodes of a massively parallel platform, is an ideal fast and low-power candidate for a local subdomain or coarse-grid solver to tackle the local components of a much larger sparse linear system.

\section{Linear Algebra on Analog Hardware}
\label{sec-analog-HW}
\begin{figure}
    \centering
    \includegraphics[width=0.48\textwidth]{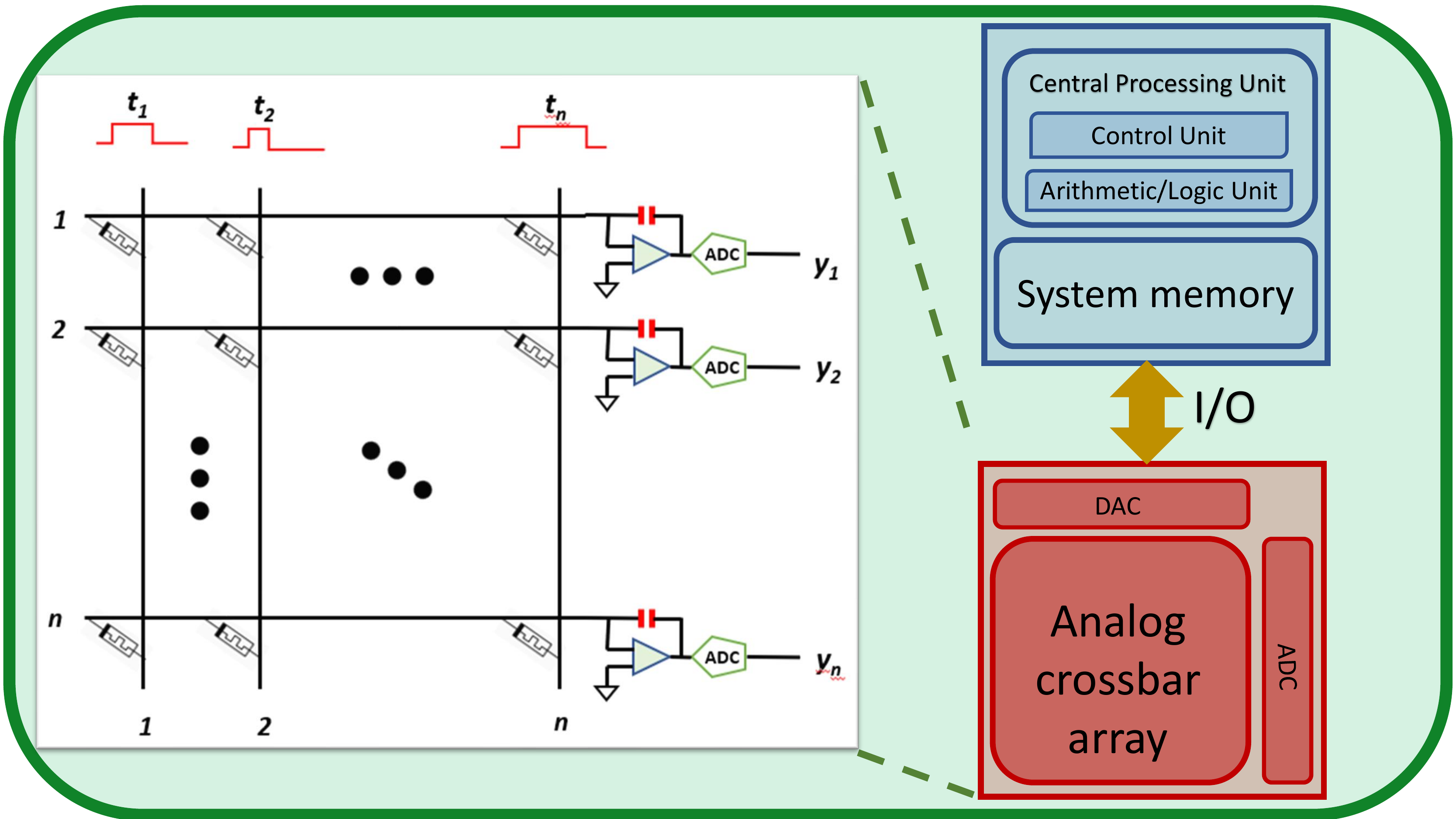}
    \caption{\textit{A hybrid digital-analog architecture consisting of a CPU and a memristive crossbar array for multiplying an $n\times n$ matrix $M$ with a vector $r$ to approximate $y = Mr$ in $O(1)$ time.
    }}
    \label{fig-RPU}
\vspace*{-0.2in}
\end{figure}

We begin by briefly describing our architecture model and its key properties. 
Figure~\ref{fig-RPU} illustrates a hybrid digital-analog architecture that combines 
a conventional microprocessor system with an analog crossbar array for performing MVM. 
The crossbar
consists of rows and columns of conductors 
with a memristive element at each row-column intersection~\cite{chua:1971}. The 
conductance of these elements can be set, reset, or updated in a non-volatile manner. 
Matrix operations are performed by mapping the $n\times n$ operand matrix 
$M$ onto the crossbar array, where the conductance $G_{ij}$ at the intersection of row $i$ and column $j$ represents the value $M[i,j]$ 
(i.e., the $(i,j)$-th entry of the matrix $M$)
after a suitable 
scaling.

The MVM operation $y = Mr$ can be emulated by sending pulses of $V_{in}$ volts along the 
columns of the crossbar
such that the length $t_j$ of the pulse along column 
$j$ is proportional to the $j$th entry $r[j]$ of vector $r$, with suitable 
normalization.
Following Ohm's Law, this contributes a current equal to $V_{in}G_{ij}$ 
on the conductor corresponding to row $i$ for a duration $t_j$. Following Kirchoff's 
Current Law, the currents along each row accumulate and can be integrated over the time period equivalent to the maximum 
pulse length using capacitors, yielding a charge proportional to $\sum_{j=1}^nM[i,j]r[j]$, 
which in turn is proportional to $y[i]$. This integrated value can be recovered as a digital quantity via an analog-to-digital converter (ADC), and yields an approximation $\widehat{y}[i]$ to $y[i]$. 

The above procedure involves multiple sources of nondeterministic noise so that the output $\widehat{y} \in \mathbb{R}^n$ is equivalent to a low precision approximation of $y$. Writing $M$ to the crossbar array incurs multiplicative and additive write noises $N^{Wm} \in \mathbb{R}^{n\times n}$ and $N^{Wa} \in \mathbb{R}^{n\times n}$, respectively, and the actual conductance values at the crosspoints of the array reflect $\widehat{M} = M\odot(I + N^{Wm}) + N^{Wa}$, where $\odot$ denotes element-wise multiplication. Similarly, digital-to-analog conversion (DAC) of the vector $r$ into voltage pulses suffers from multiplicative and additive input noises $N^{Im} \in \mathbb{R}^n$ and $N^{Ia} \in \mathbb{R}^n$, respectively. As a result, the matrix $\widehat{M}$ is effectively multiplied by a perturbed version of $r$ given by $\widehat{r} = r\odot(\mathbf{1} + N^{Im}) + N^{Ia}$, where $\mathbf{1}$ is a vector of all ones.

A characteristic equation to describe the output $\widehat{y}$ of an analog MVM $Mr$ can be written as
\begin{equation}
\label{eqn-error1}
\widehat{y} = \widehat{M}\widehat{r} \odot (\mathbf{1} + N^{Om}) + N^{Oa},
\end{equation}
where $N^{Om} \in \mathbb{R}^n$ and $N^{Oa} \in \mathbb{R}^n$ denote the multiplicative and additive components of the output noise, respectively. These components reflect the inherent inexactness in the multiplication based on circuit laws and current integration, as well as the loss of precision in the ADC conversion of the result vector. 
Equation~\eqref{eqn-error1} can be simplified as
\begin{equation}
\label{eqn-error2}
\widehat{y} = (M+E)r + \widehat{N}^{Oa},
\end{equation}
where $\widehat{N}^{Oa}$ is the overall effective additive noise that captures all the lower order noise terms from Equation~\eqref{eqn-error1} and the nondetermistic error matrix $E$ is given by
\begin{equation}
\label{eqn-error3}
E = M\odot(N^{Wm}+\mathbf{1}\otimes N^{Im} + N^{Om}\otimes\mathbf{1}) + N^{Wa};
\end{equation}
here $\otimes$ denotes outer product. We provide additional technical details in Appendix~\ref{app:IR}.

The exact characteristics and magnitudes of the noises depend on the physical implementation~\cite{Haensch2019,LeGallo2018} of the analog array and the materials used therein, the details of which are beyond the scope 
of this paper but they are generally captured by our model. Regardless of the implementation, the noises are always device dependent and stochastic. 

Once the matrix $M$ is written to the crossbar array, analog MVMs involving the matrix can be performed in $O(1)$ time, resulting in an $O(n^2)$ speedup over 
the equivalent digital computation if $M$ is dense. Therefore, a linear system solver that can overcome the analog MVM error, even if it requires many more MVM operations than its digital counterpart, 
can substantially reduce the cost of the solution. 

\section{An iterative solver based on memristive approximate inverse preconditioning}
\label{sec-flexsolvers}

The goal of preconditioning is to transform an $n\times n$ system of linear equations $Ax=b$ to enable an iterative procedure to compute a sufficient approximation of $x$ in fewer steps. Algebraically, preconditioning results 
in solving the system $MAx=Mb$, where $M\in \mathbb{R}^{n\times n}$ is an approximation of $A^{-1}$. 
Since $A^{-1}$ is dense in general, most popular general-purpose preconditioners 
are applied implicitly; 
e.g.,
the ILU preconditioner generates a lower (upper) triangular 
matrix $L$ ($U$) such that $A\approx LU$. The preconditioner $M=U^{-1}L^{-1}$ is 
then applied by forward/backward  
substitution~\cite{saad2003iterative}. 

\subsection{Approximate inverse matrices as preconditioners}

Although the inverse of a sparse matrix is dense in general, a significant fraction of the entries in the inverse matrix often have small magnitudes~\cite{benzi1999bounds,demko1984decay}.
Approximate inverse preconditioners~\cite{benzi1999comparative,chow2001parallel,grote1997parallel} exploit this fact and seek to compute
an approximation $M$ to $A^{-1}$ that can be applied via 
an MVM operation.  
Computing $M$ can be framed as an optimization 
problem that seeks to minimize $||I-MA||_F$. 
One popular method for computing $M$ is by the Sparse Approximate Inverse (SPAI) technique~\cite{grote1997parallel}, which updates the sparsity pattern of $M$ 
dynamically by repeatedly choosing new profitable indices for each column of $M$ 
that lead to an improved approximation of $A^{-1}$. The quality of the approximation 
is controlled by computing the norm of the residual $I(:,j)-AM(:,j)$ after each 
column update. The $j$th column of $M$ has at most 
$\mathtt{nnz}_{\mathrm{AI}}$ nonzero entries and satisfies 
$\|AM(:,j)-I(:,j)\|_F\leq \mathtt{tol}_{\mathrm{AI}}$, for a given threshold 
tolerance $\mathtt{tol}_{\mathrm{AI}} \in \mathbb{R}$. 

An effective approximate inverse preconditioner $M$ can still be considerably denser than the coefficient matrix $A$. Therefore, preconditioning must lead to a drastic 
convergence improvement for a reduction in the overall wall-clock time. 
Applying approximate inverse preconditioners as proposed in this paper has the advantage that the computation time and energy consumption of preconditioning would be much lower on an analog crossbar array than on digital hardware, even at low precisions. Since MVM with $M$ can be performed in $O(1)$ time on an analog array regardless of the number of nonzeros in $M$, denser approximate inverses may be utilized without increasing the associated costs. 

While analog crossbar arrays can speed up the application of the approximate 
inverse preconditioner, the accuracy of the application 
step is constrained by the various nondeterministic 
noises in the analog hardware discussed in Section~\ref{sec-analog-HW}. This 
restricts the use of popular Krylov subspace approaches such as 
GMRES~\cite{saad1986gmres}, which require a deterministic 
preconditioner. Although flexible variants~\cite{saad1993flexible} exist, 
in which the preconditioner can vary from one step to another, these 
variants generally require the orthonormalization of large subspaces and 
perform a relatively larger number of Floating Point Operations (FLOPs) 
outside the preconditioning step. Therefore, in this paper, we consider a stationary iterative scheme.

\subsection{Preconditioned Richardson iteration}

Stationary methods approximate the solution of a linear system $Ax=b$ by iteratively applying an update of the form $x_{i}=Dx_{i-1}+d$, where $D\in \mathbb{R}^{n\times n}$ and $d\in \mathbb{R}^n$ are fixed.
Preconditioned Richardson iteration can be defined as
\begin{equation*}
    \begin{aligned}
    \alpha MAx &=\alpha Mb,\ \ \ \alpha \in \mathbb{R}, \\
    x &= x + \alpha Mb-\alpha MAx\\
    &=(I-\alpha MA)x + \alpha Mb, \\
    x_i&=(I-\alpha MA)x_{i-1} + \alpha Mb,
    \end{aligned}
\end{equation*}
which is identical to the form $x_{i}=Dx_{i-1}+d$, with $D=I-\alpha MA$ and 
$d=\alpha Mb$ \cite{richardson1911ix}. 
The approximate solution $x_i$ produced by preconditioned Richardson iteration 
during the $i$th iteration satisfies
\begin{equation*}
    \begin{aligned}
    x_i-x & = x_{i-1}-x-\alpha MAx_{i-1} + \alpha MAx \\ 
    & = (x_{i-1}-x)-\alpha MA(x_{i-1}-x) \\ 
    & = (I-\alpha MA)(x_{i-1}-x) , \\
    \|x_i-x\| & \leq \|I-\alpha MA\|\|x_{i-1}-x\|\\ 
    &\leq \left(\|I-\alpha MA\|\right)^i\|x_{0}-x\| ,
    \end{aligned}
\end{equation*}
where the latter inequality holds 
for any vector norm and the corresponding induced matrix norm. A similar bound can also be shown for the residual vector $r_i=b-Ax_{i}$; i.e.,
\begin{equation*}
    \|r_i\| \leq \left(\|I-\alpha AM\|\right)^i\|r_0\|.
\end{equation*}
The above inequalities show that the sequence $\lim\limits_{i\rightarrow \infty}x_i$ 
will converge to the true solution $x$ as long 
as the preconditioner $M$ and scaling scalar $\alpha$ are chosen such that 
$\|I-\alpha MA\|<1$. In fact, the limit of the iterative process will converge 
to $x$ as long as the spectral radius $\rho(I-\alpha MA)$ of the matrix $I-\alpha MA$ is 
strictly less than one. Nonetheless, we use the spectral norm since $\rho(I-\alpha MA) 
\leq \|I-\alpha MA\|$. If we denote the eigenvalues of the matrix $MA$ by $|\lambda_1|
\leq \ldots\leq |\lambda_n|$, then we need to pick $\alpha$ such that $|1-\alpha \lambda_n|<1$, 
i.e., we need
$0< \alpha < \dfrac{2\lambda_n^*}{|\lambda_n|^2}$.
We use $\alpha=1$ in the remainder of the paper.

\subsection{Preconditioning through memristive hardware}

Algorithm \ref{alg:richardson} summarizes our hybrid Richardson iterations 
preconditioned by applying an approximate inverse through analog crossbar 
hardware. The iterative procedure terminates when either the maximum 
number of iterations $m_{\mathrm{it}}$ is reached or the relative residual 
norm drops below a given threshold tolerance $t_{\mathrm{ol}}\in \mathbb{R}$. 
Each iteration of Algorithm~\ref{alg:richardson} requires a sparse MVM with 
the coefficient matrix $A$ (Line 5), two ``scalar alpha times x plus y'' (AXPY)~\cite{Lawson:1979:BLA} operations (Lines 5 and 7), a vector norm 
computation (Line 6), and an MVM with the approximate inverse preconditioner 
$M$ (Line 7) computed in analog hardware. Therefore, of the $3n + 2(\mathtt{nnz}(A)+\mathtt{nnz}(M))$ total FLOPs per iteration, $2\mathtt{nnz}(M)$ 
can be performed in $O(1)$ time on the analog hardware.

\begin{algorithm}
    \caption{Richardson iterations with approximate inverse preconditioning on hybrid hardware}
    \label{alg:richardson}
    \begin{algorithmic}[1]
        \STATE {\bf input}: $A\in \mathbb{R}^{n\times n};\ b,x_0\in \mathbb{R}^n;\ t_{\mathrm{ol}}\in \mathbb{R};\ m_{\mathrm{it}}\in \mathbb{N}$.
        \STATE Construct approximate inverse preconditioner $M \in \mathbb{R}^{n\times n}$;
        \STATE Write $M$ to the analog crossbar array;
        \FOR {$i=0$ \TO $m_{\mathrm{it}}-1$}
            \STATE $r_i = b - Ax_i$; // Digital MVM and AXPY
            \STATE If ($\|r_i\|_2 \leq t_{\mathrm{ol}} \|b\|$) exit;
            \STATE $x_{i+1} = x_i + Mr_i$; // Analog MVM and digital AXPY
        \ENDFOR
        \RETURN $x_i$;
    \end{algorithmic}
\end{algorithm}

The ideal per-iteration speedup that the hybrid implementation can achieve over its digital counterpart is 
    ${\cal S}_{\mathrm{ideal}}=1+\dfrac{2\mathtt{nnz}(M)}{3n+2\mathtt{nnz}(A)}$,
which increases with the density $\mathtt{nnz}(M)$ of the preconditioner. 
In reality, the hybrid implementation is likely to require more iterations to reach the
same residual norm as the digital implementation because the noise introduced by the analog hardware leads to a less accurate 
application of the preconditioner $M$. More specifically, excluding the time spent on constructing $M$, the overall 
computational speedup achieved by the hybrid implementation is 
\begin{equation*} \label{speedup2}
{\cal S}_{\mathrm{tot}}=\dfrac{m_{\mathrm{d}}}{m_{\mathrm{h}}}\left(1+\dfrac{2\mathtt{nnz}(M)}{3n+2\mathtt{nnz}(A)}\right),
\end{equation*}
where $m_{\mathrm{d}}$ and $m_{\mathrm{h}}$ denote 
the number of digital and hybrid preconditioned Richardson 
iterations, respectively.

\subsection{The effects of device noise}
\label{sec-noise-analysis}
Since the MVM between the preconditioner $M$ and the residual vector of the $i$th iteration $r_{i}$ is computed through an analog crossbar array, the 
inherent inaccuracies of the analog device lead to an update of the form 
$$x_{i+1} = x_{i} + (M+E)r_{i} + \widehat{N}^{Oa}$$ from Equation~\eqref{eqn-error2}. Note that $E$ and $\widehat{N}^{Oa}$ stochastically and nondeterministically vary between iterations.
The error norm $\|x_{i+1}-x\|$ then satisfies the
relationships
\begin{equation*}
    \|x_{i+1}-x\| \geq \|\widehat{N}^{Oa}\|
\end{equation*}
and
\begin{equation*}
    \|x_{i+1}-x\| \leq \|I-(M+E)A\|\|x_{i}-x\|.
\end{equation*}

The additive noise $\widehat{N}^{Oa}$ has a direct impact on the lower bound of the
error that can be achieved. Additionally, the norm of the error $\|x_{i+1}-x\|$ will be smaller than $\|x_{i}-x\|$ as 
long as $\|I-(M+E)A\| < 1$,
for which we have the upper bound
\begin{equation*}
    \|I-(M+E)A\|\leq \|I-MA\|+\|E\|\|A\|.
\end{equation*}
Under the assumption that the spectral norm of the matrix $I-MA$ is less than one, 
i.e., $\|I-MA\| <1$, a sufficient condition for the spectral norm of the matrix 
$I-(M+E)A$ to be less than one is 
\begin{equation*}
    \|E\|<\dfrac{1-\|I-MA\|}{\|A\|}.
\end{equation*}
Additionally, for $\|I-(M+E)A\|<1$ to hold under the condition $M=A^{-1}+\Delta$, $\|E\| < \|A\|^{-1}-\|M-A^{-1}\| $.

Consistent with the original deterministic analysis for iterative refinement by Moler~\cite{Moler} from a digital perspective, but adapted to Richardson iteration with noisy analog devices, a sufficiently small $\|E\|$ in a stochastic sense guarantees the convergence of the preconditioned Richardson iteration process.
In particular, we need to have $\|{E}\| < \frac{1-\|I-MA\|}{\|A\|}$ hold in a sufficiently large number of iterations.
To understand and ensure when these conditions will hold, we establish explicit bounds for $\|E\|$ in terms of the statistical characteristics of the elements of the matrix $E$. We omit the details here and refer the interested reader to
Appendix~\ref{app:IR}.

\section{Simulation Experiments} 
\label{sec-experiments}

Our experiments were conducted in a Matlab environment (version R2020b) with 64-bit 
arithmetic on a single core of a 2.3 GHz 8-Core 
Intel i9 machine with 64 GB of system memory. We used a Matlab version of the publicly available simulator~\cite{rasch2021flexible} with a PyTorch interface for emulating the noise, timing, and energy characteristics of an analog crossbar array. The simulator models all sources of analog noise outlined 
in Section~\ref{sec-analog-HW} as scaled Gaussian processes. Using Matlab notation, 
the components of the matrix write noise were modeled as $\mathtt{randn(\cdot)\times 5.0\mathtt{e}-3}$, and those of the input and output noises were both modeled as $\mathtt{randn(\cdot)\times 1.0\mathtt{e}-2}$; these are the default settings in the simulator based on currently realizable analog hardware~\cite{Gokmen2016}. 
The number of bits used in the ADC and DAC was set to $7$ and $9$, respectively. Table~\ref{tab1} shows the default parameters used throughout our experiments to simulate 
the analog device and to construct the approximate inverse preconditioner. 

\begin{table}[ht]
\centering
\caption{\textit{Default parameters.}} \label{tab1}
\begin{tabular}[t]{llcc}
\toprule
Module& Parameter & Value\\
\midrule
Analog device& $N^W$&$5.0\mathtt{e}-3$\\
Analog device& $N^I$ &$1.0\mathtt{e}-2$\\
Analog device& $N^O$&$1.0\mathtt{e}-2$\\\hline
Richardson it. &$t_{\mathrm{ol}}$ & $1.0\mathtt{e}-5$\\
Richardson it. &$m_{\mathrm{it}}$ & $50$\\\hline 
Approximate inverse&$\mathtt{nnz}_{\mathrm{AI}}$ & $40\times \mathtt{nnz}(A)$\\
Approximate inverse&$\mathtt{tol}_{\mathrm{AI}}$ & $5.0\mathtt{e}-2$\\
\bottomrule
\end{tabular}
\end{table}%


\begin{figure}
    \centering
    \includegraphics[width=0.24\textwidth]{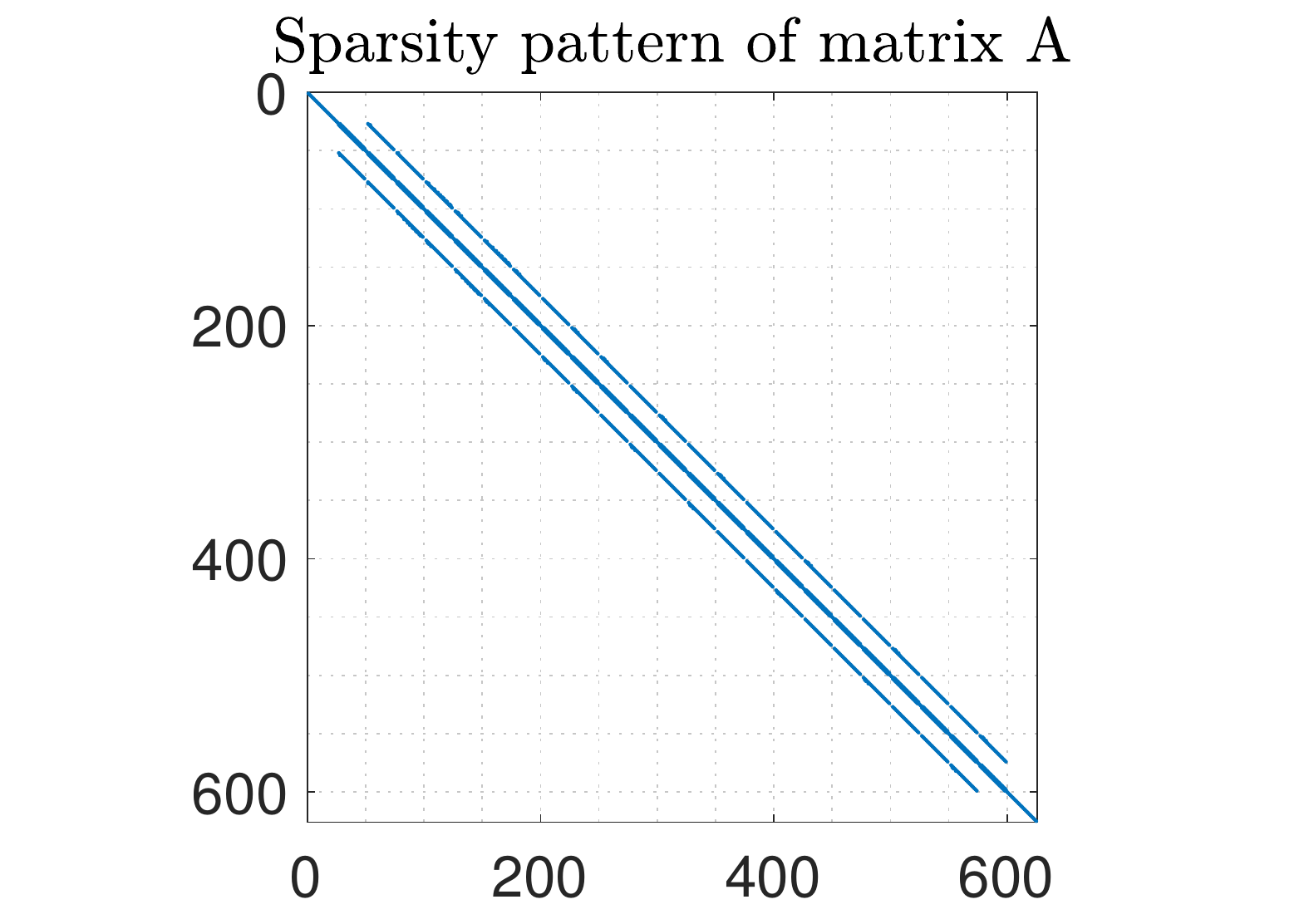}
    \includegraphics[width=0.24\textwidth]{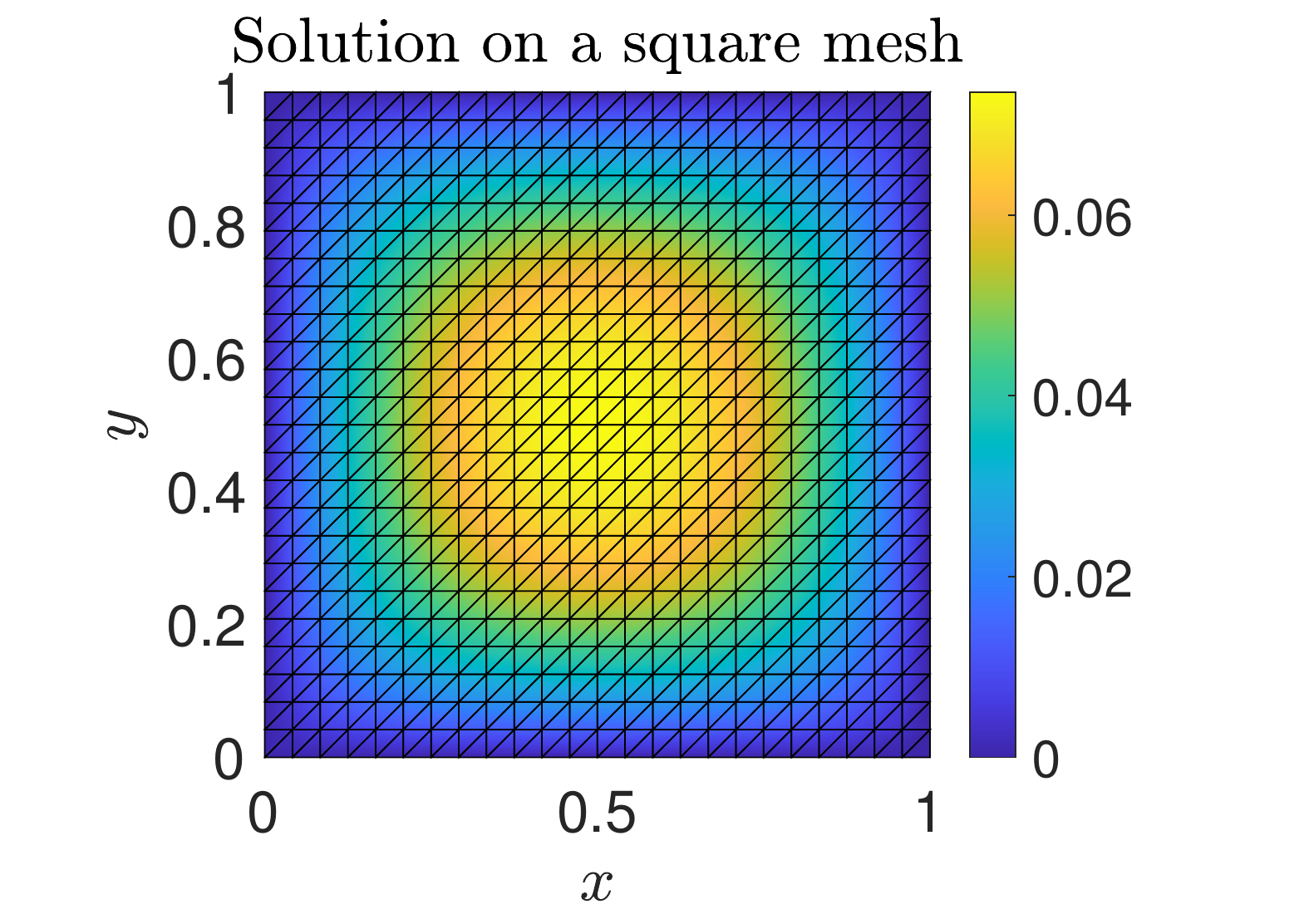}
    \includegraphics[width=0.24\textwidth]{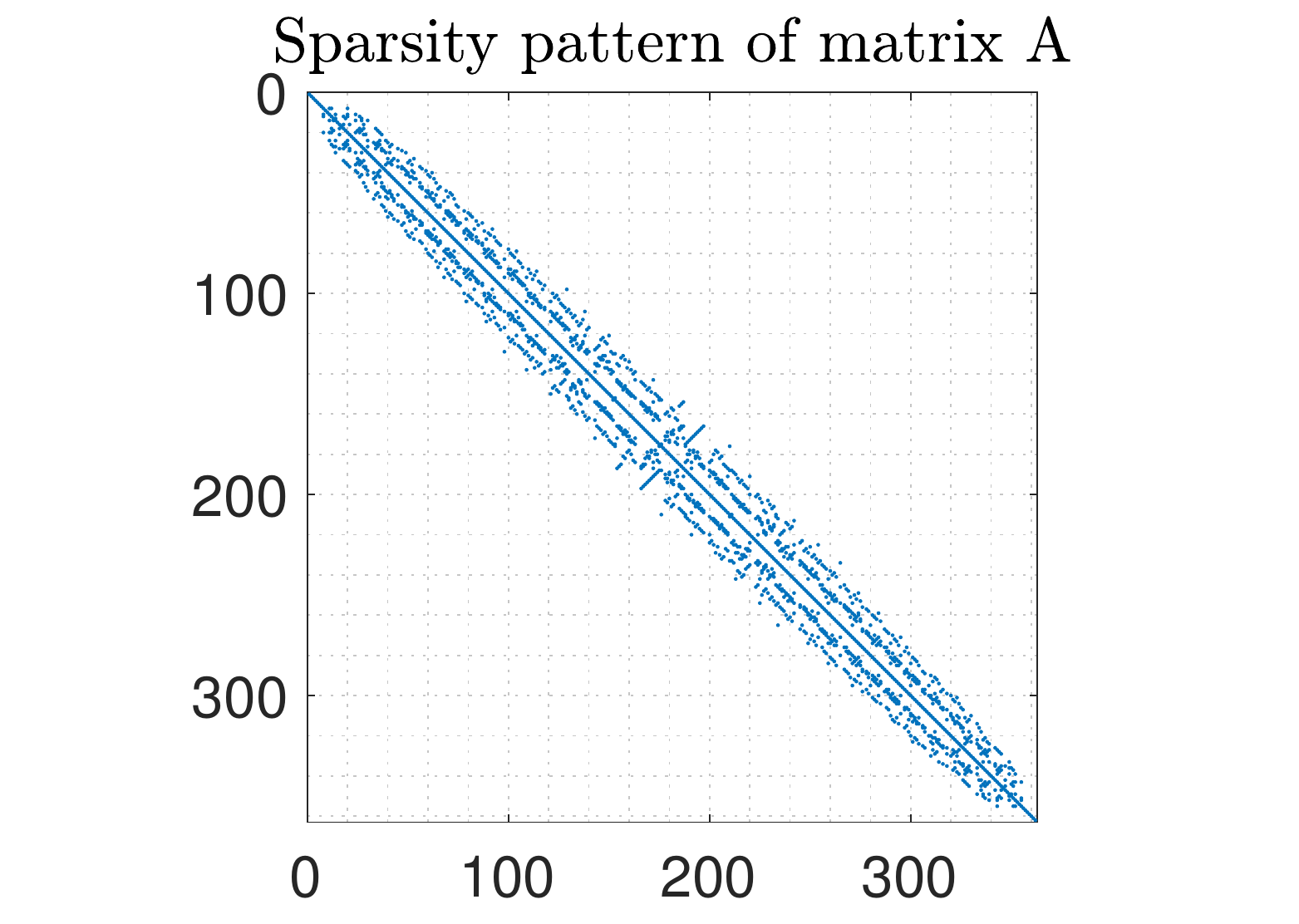}
    \includegraphics[width=0.24\textwidth]{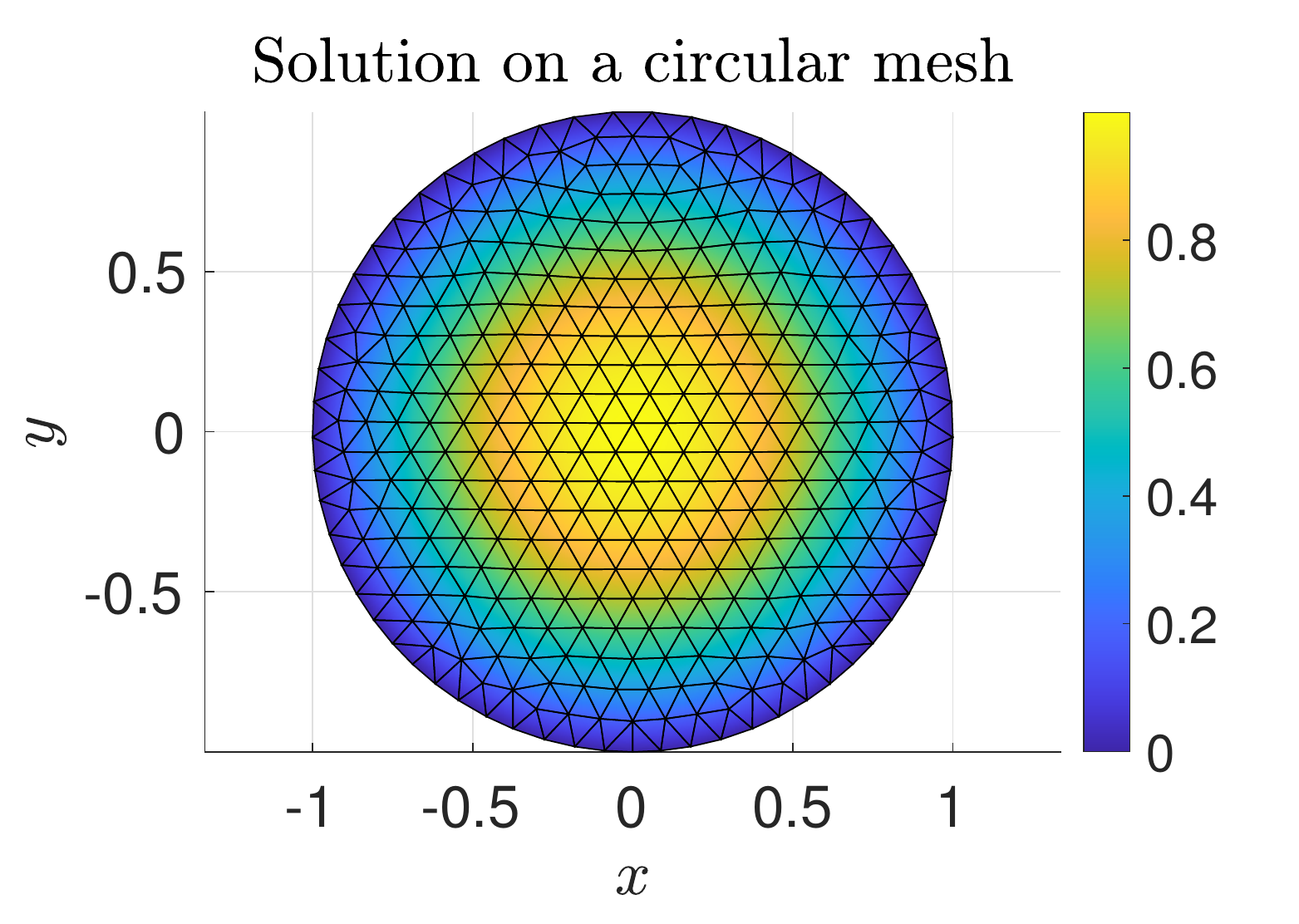}
\caption{\textit{Left: Sparsity patterns of the discretized Laplacian for the square mesh $\Omega:=[0,1]^2$ (top) and the circular mesh $\Omega:= \{x^2+y^2=1\}$ (bottom). Right: Surface plots of the solutions $u(x,y)$.  }} \label{spy2}
    \vspace*{-0.2in}
\end{figure}

Our test problems originate from discretizations of model PDEs. The first two matrices are derived from a Finite Element (FE) discretization of Poisson's equation on $\Omega \subset \mathbb{R}^2$ with homogeneous Dirichlet boundary 
conditions
\begin{equation}
    -\left(\dfrac{\partial^2}{\partial x^2}+\dfrac{\partial^2}{\partial y^2}\right)
    u(x,y) = f(x,y),\ \ u|_{\partial \Omega}=0,
\end{equation}
where $\partial$ denotes the partial derivative with respect to an independent 
variable and $f(x,y)$ denotes the load (or source) vector. 
We consider two different domains: a square domain $\Omega:=[0,1]^2$ and 
a circular domain $\Omega=\{[x,y]\in \mathbb{R}^2 \; \big| \; x^2+y^2=1\}$ of radius one centered at the origin. The Laplace operator 
$\left(\dfrac{\partial^2}{\partial x^2}+\dfrac{\partial^2}{\partial y^2}\right)u(x,y)$
is discretized by linear finite elements while the load vector is set equal to the 
constant source function $f(x,y)\equiv 1$. Figure~\ref{spy2} plots the 
sparsity patterns of the discretized Laplacian matrices $A$ and the surfaces 
of the solutions $u(x,y)$
for the square ($n=625$) and circular ($n=362$) meshes. Our third test matrix stems from a Finite Difference (FD) discretization of the Laplace operator in the unit cube with 
homogeneous Dirichlet boundary conditions and $n=8^3$.
The average number of nonzero 
entries per row in the discretized sparse matrices $A$ and in the corresponding preconditioners $M$ are listed in Table~\ref{tab2}, which shows that the fill-in factor
(i.e., the ratio 
$\mathrm{nnz}(M)/\mathrm{nnz}(A)$)
ranges
from $14$ to $24$. Our test matrices are relatively small because we are limited by the speed of the Matlab version of the simulator for the analog crossbar arrays; however, our results are meaningful because the relative advantage of analog hardware,
in general, only increases with larger matrices.

\begin{table*}[t]
\centering
\setlength{\tabcolsep}{2pt}
\caption{\textit{Comparison of Richardson iterations with digital ($M_d$) and hybrid ($M_h$) preconditioning; $m$, $m_{\mathrm{d}}$ and
$m_{\mathrm{h}}$ denote the number of non-preconditioned,
digital, and hybrid preconditioned Richardson iterations, respectively; $\kappa$ denotes the condition number of $A$.}}
\label{tab2}
\begin{tabular}[t]{l|cccc|ccc|ccc|cc}
\toprule
Problem & $n$ & $\kappa$ & $\dfrac{\mathrm{nnz}(A)}{n}$ & $\dfrac{\mathrm{nnz}(M)}{n}$ & $\rho(I-A)$ & $\rho(I-M_{\mathrm{d}}A)$  & $\rho(I-M_{\mathrm{h}}A)$ & $m$ & $m_{\mathrm{d}}$  & $m_{\mathrm{h}}$
& FLOPs($M_{\mathrm{d}}$)  & FLOPs($M_{\mathrm{h}}$)\\
\midrule
FE Square  & 625 & $3.4\mathtt{e}+2$ & 4.4 & 93.5 & 0.99 & 0.75 & 0.75 & Failed & 41 & 44 & $5.0\mathtt{e}+6$ & $3.1\mathtt{e}+5$ \\
FE Circular  & 362 & $2.0\mathtt{e}+2$ & 5.6 & 86.6 & 0.97 & 0.55 & 0.55 & Failed & 21 & 23 & $1.4\mathtt{e}+6$ & $1.1\mathtt{e}+5$\\
FD 3D & 512 & $6.2\mathtt{e}+1$ & 6.2 & 81.1 & 0.75 & 0.17 & 0.54 & Failed & 7 & 16 & $6.3\mathtt{e}+5$ & $1.2\mathtt{e}+5$ \\
\bottomrule
\end{tabular}
\end{table*}

\begin{figure*}
    \centering
    \includegraphics[width=0.325\textwidth]{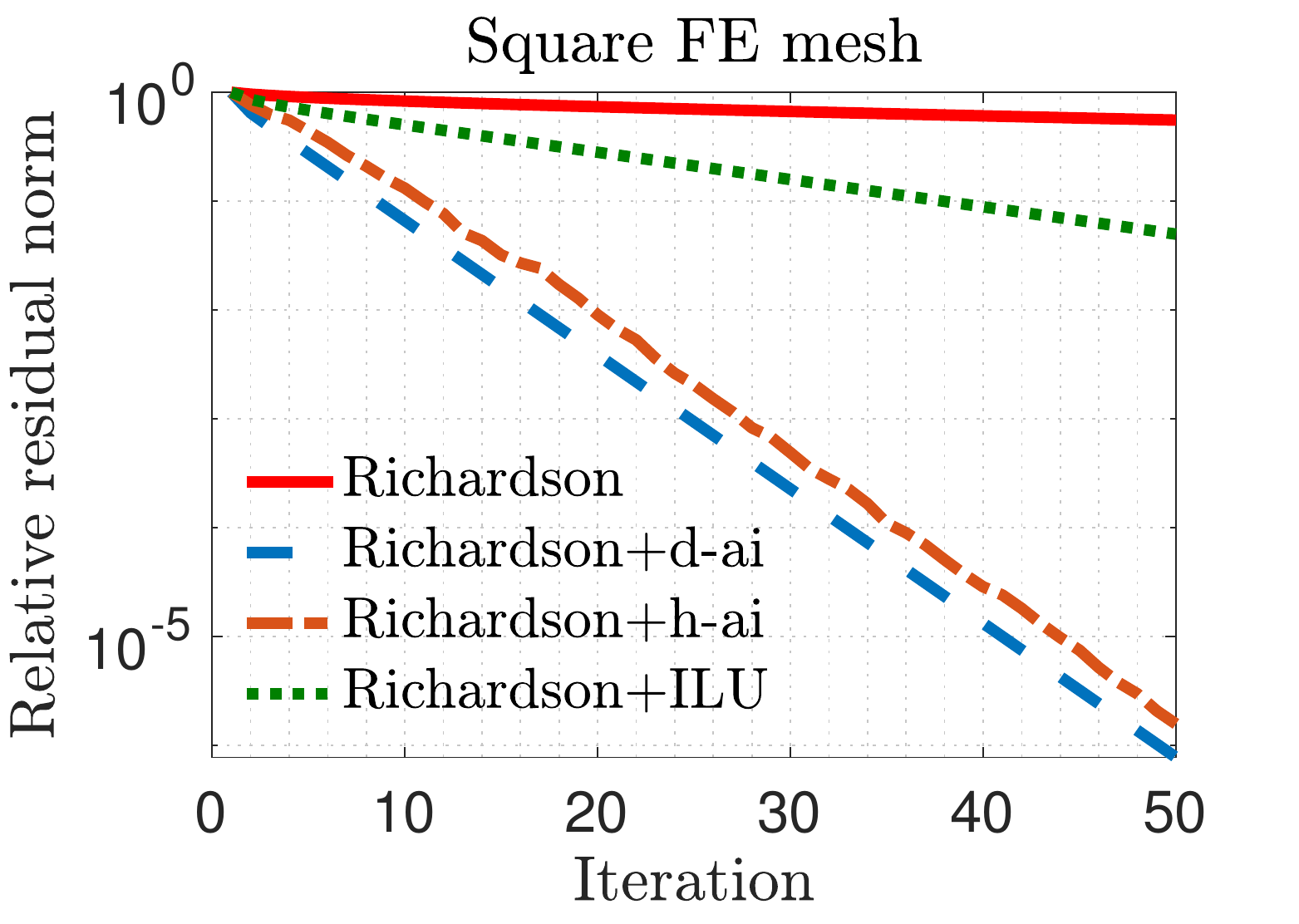}
    \includegraphics[width=0.325\textwidth]{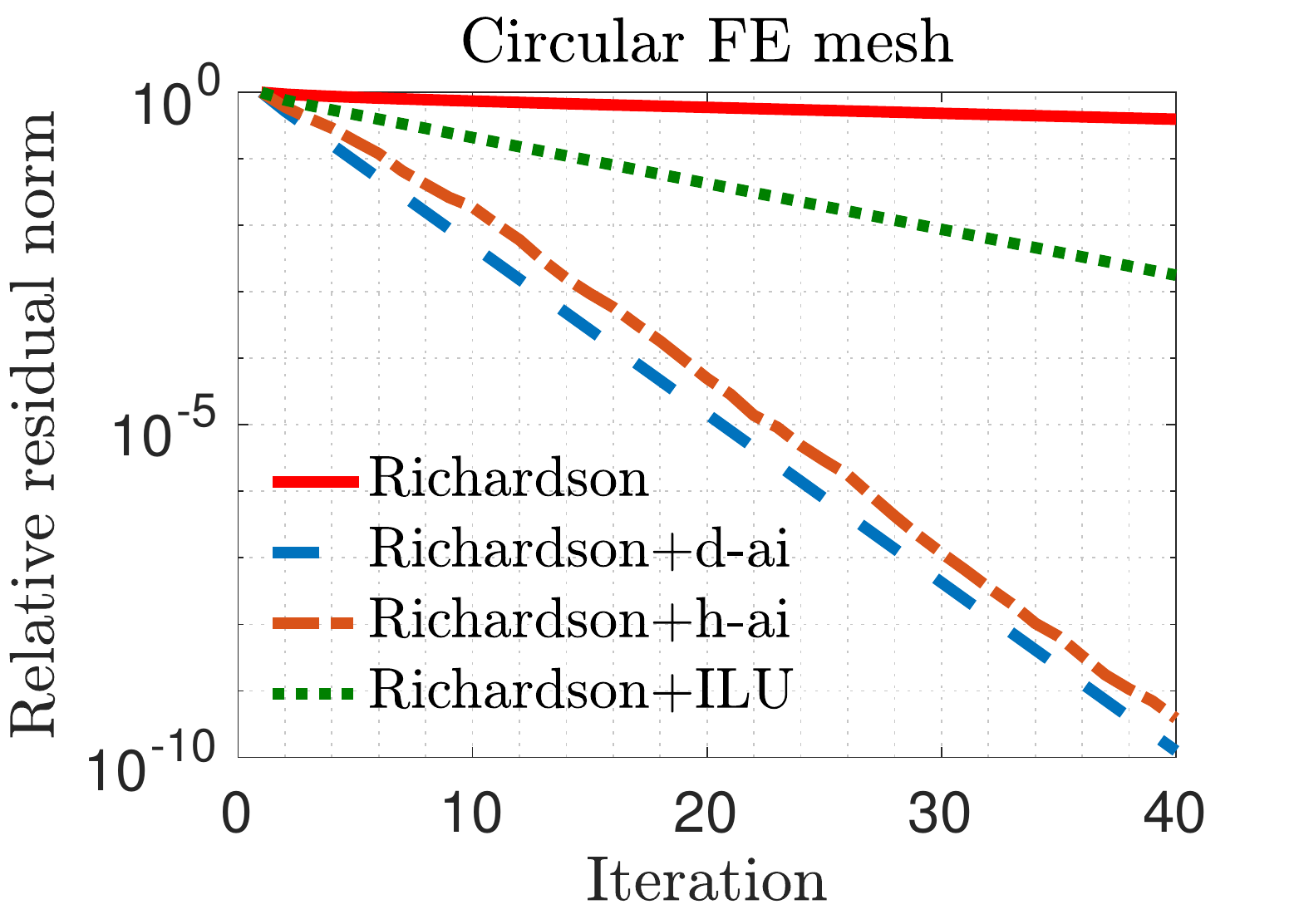}
    \includegraphics[width=0.325\textwidth]{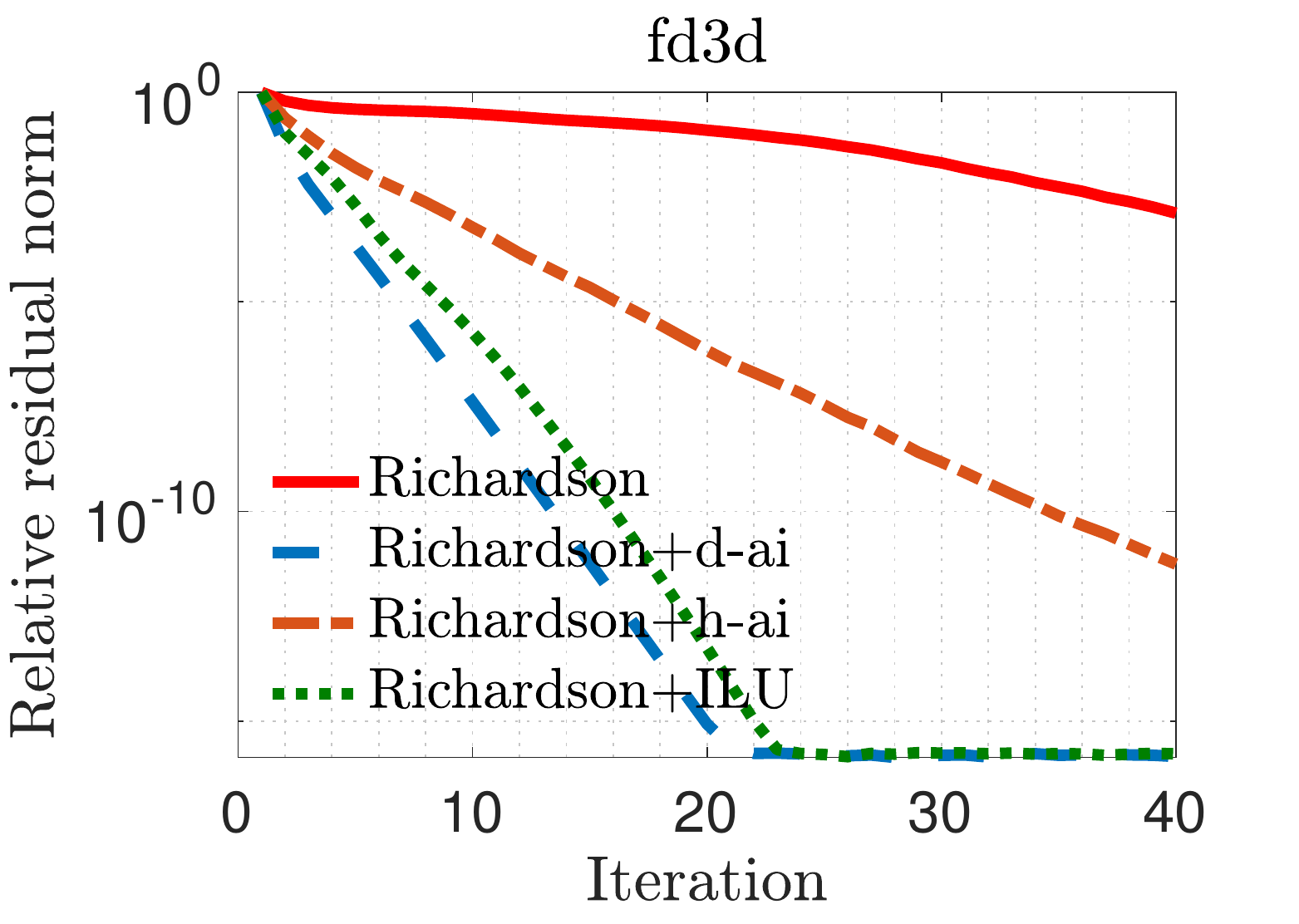}
    \caption{\textit{Relative residual norm achieved by Richardson iterations without and with digital/hybrid approximate inverse preconditioning}.} \label{fig1}
\vspace*{-0.1in}
\end{figure*}

\begin{figure*}
    \centering
    \includegraphics[width=0.325\textwidth]{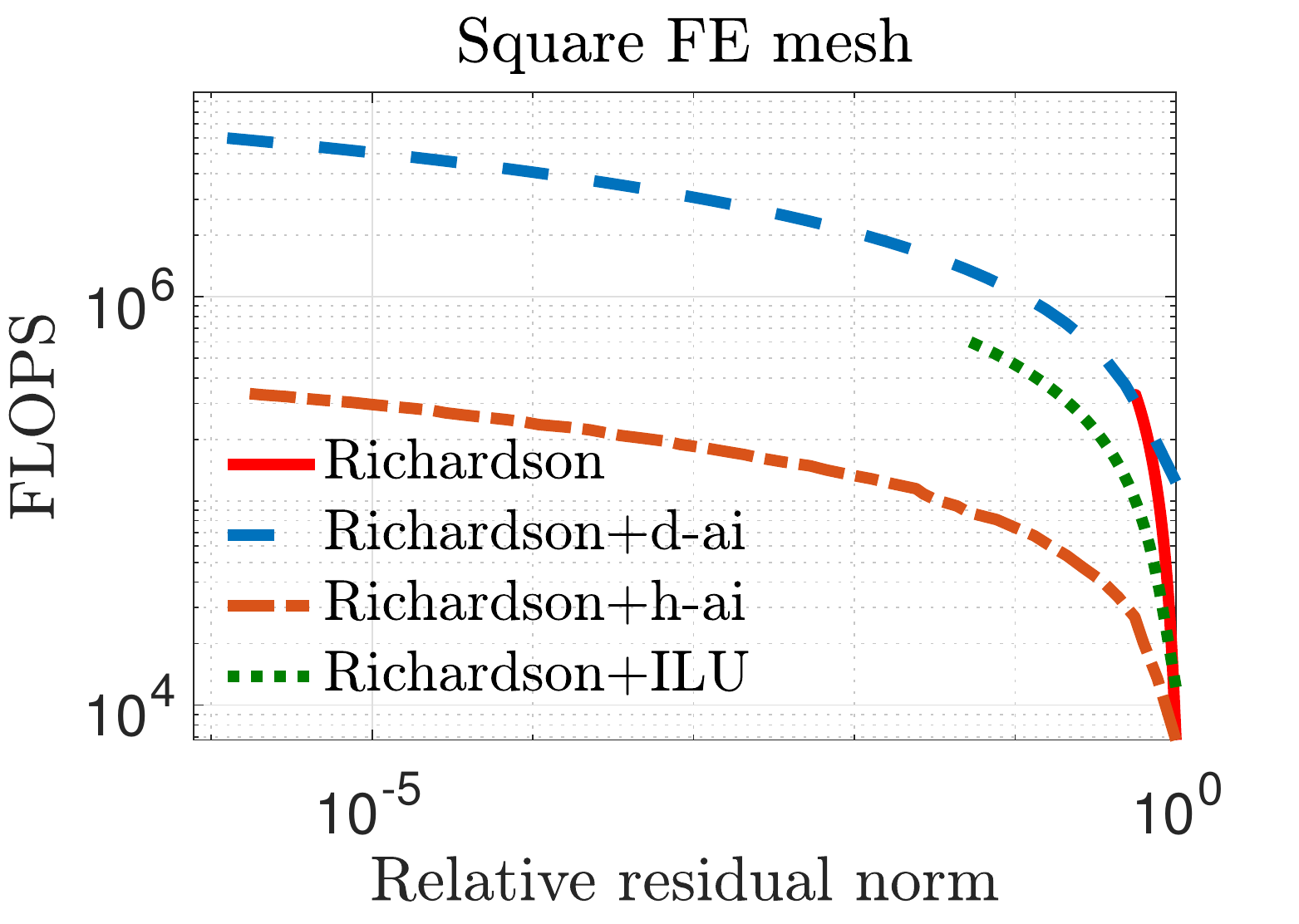}
    \includegraphics[width=0.325\textwidth]{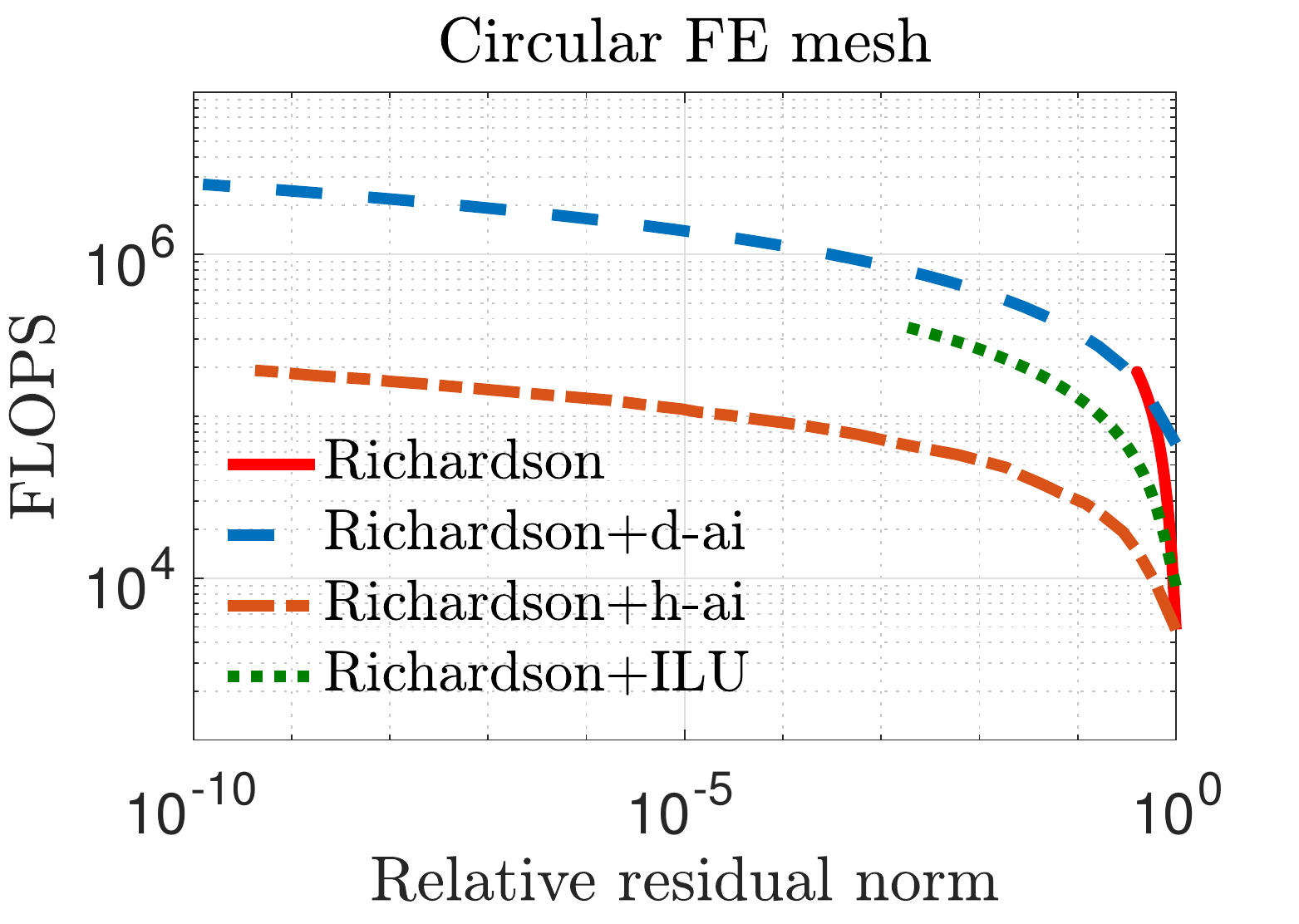}
    \includegraphics[width=0.325\textwidth]{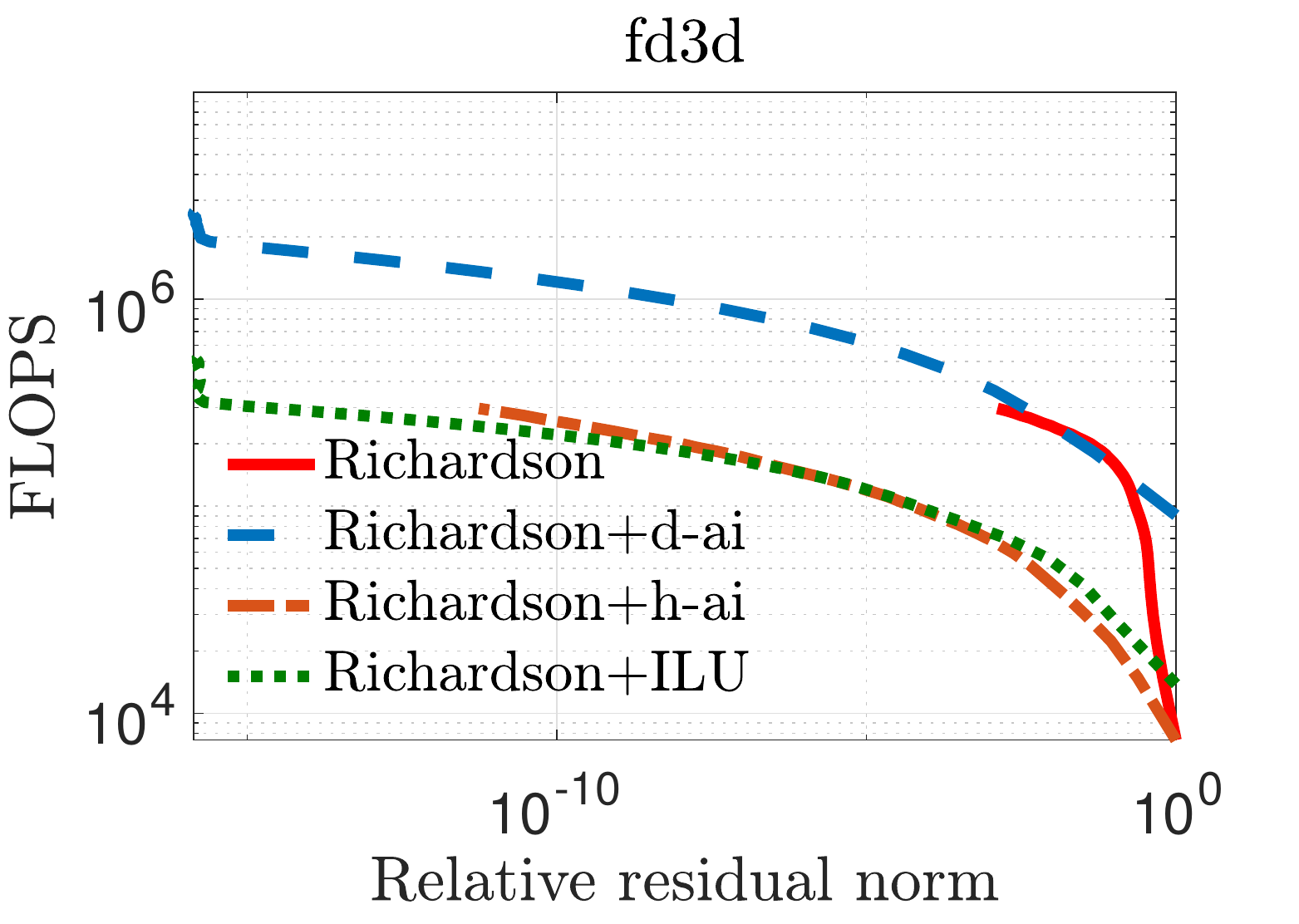}
    \caption{\textit{FLOPs as a function of the achieved relative residual norm for each of the three problems in Table~\ref{tab2}}.} \label{fig2}
\vspace*{-0.1in}
\end{figure*}

\begin{figure*}
    \centering
    \includegraphics[width=0.315\textwidth]{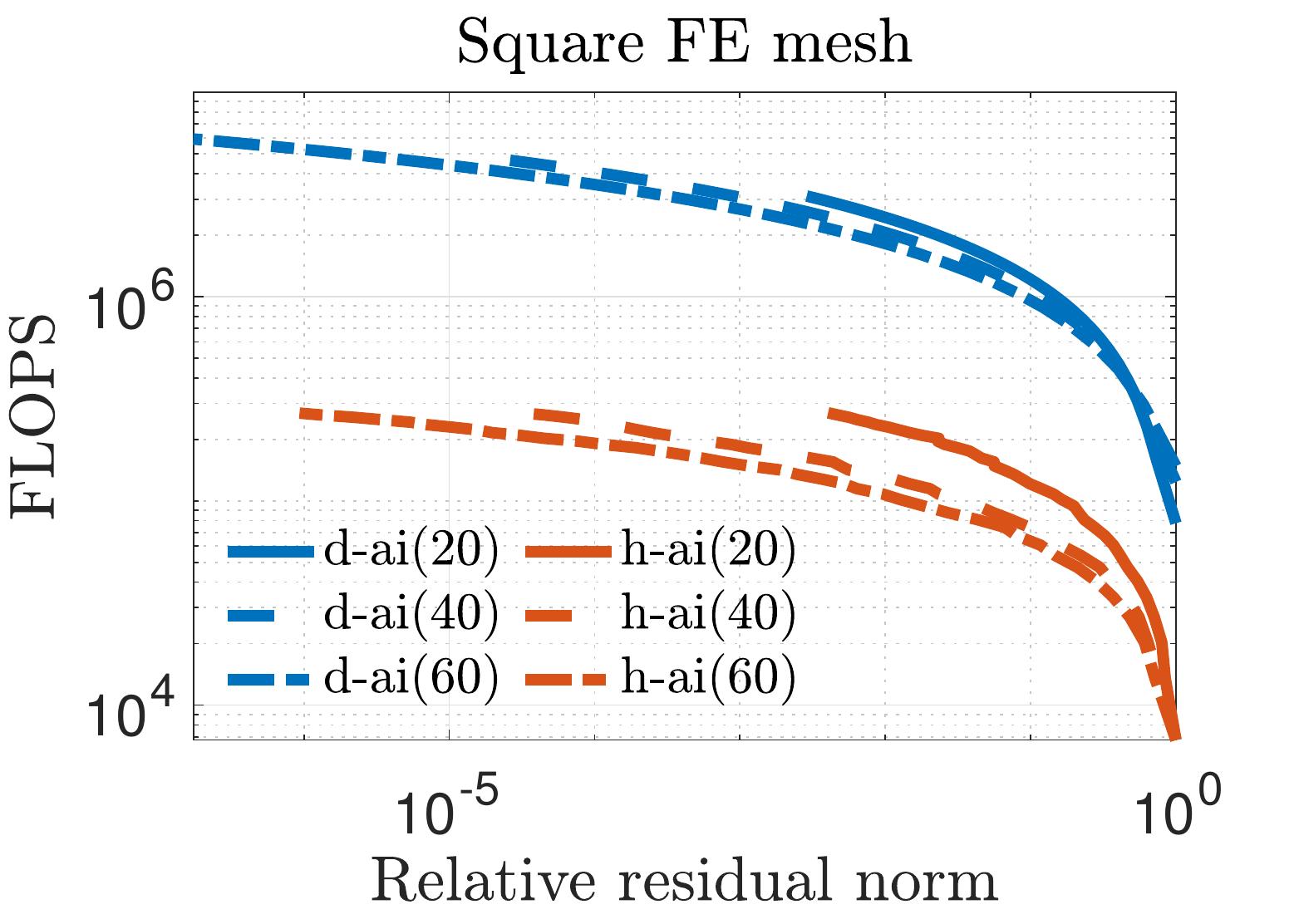}
    \includegraphics[width=0.315\textwidth]{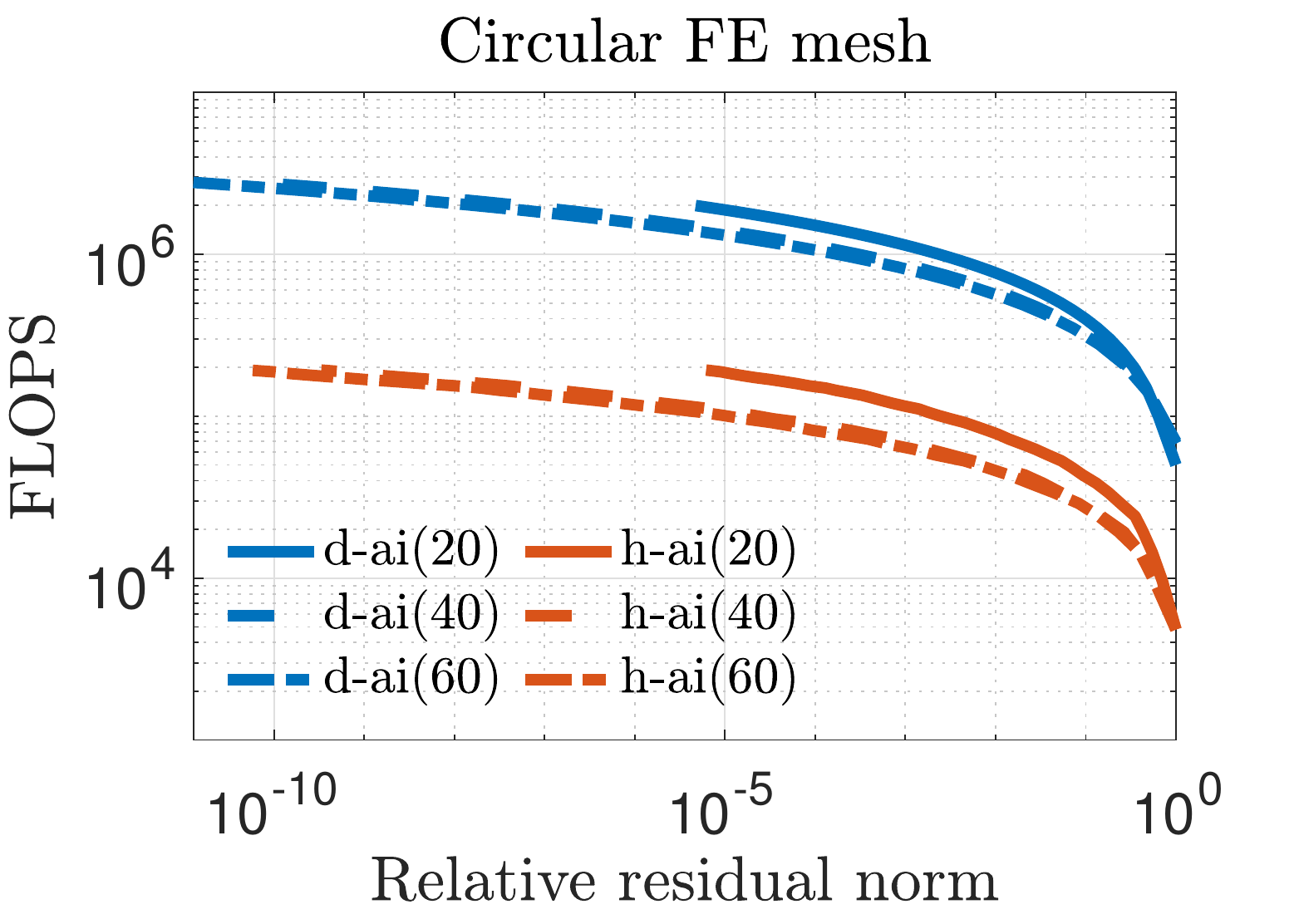}
    \includegraphics[width=0.34\textwidth]{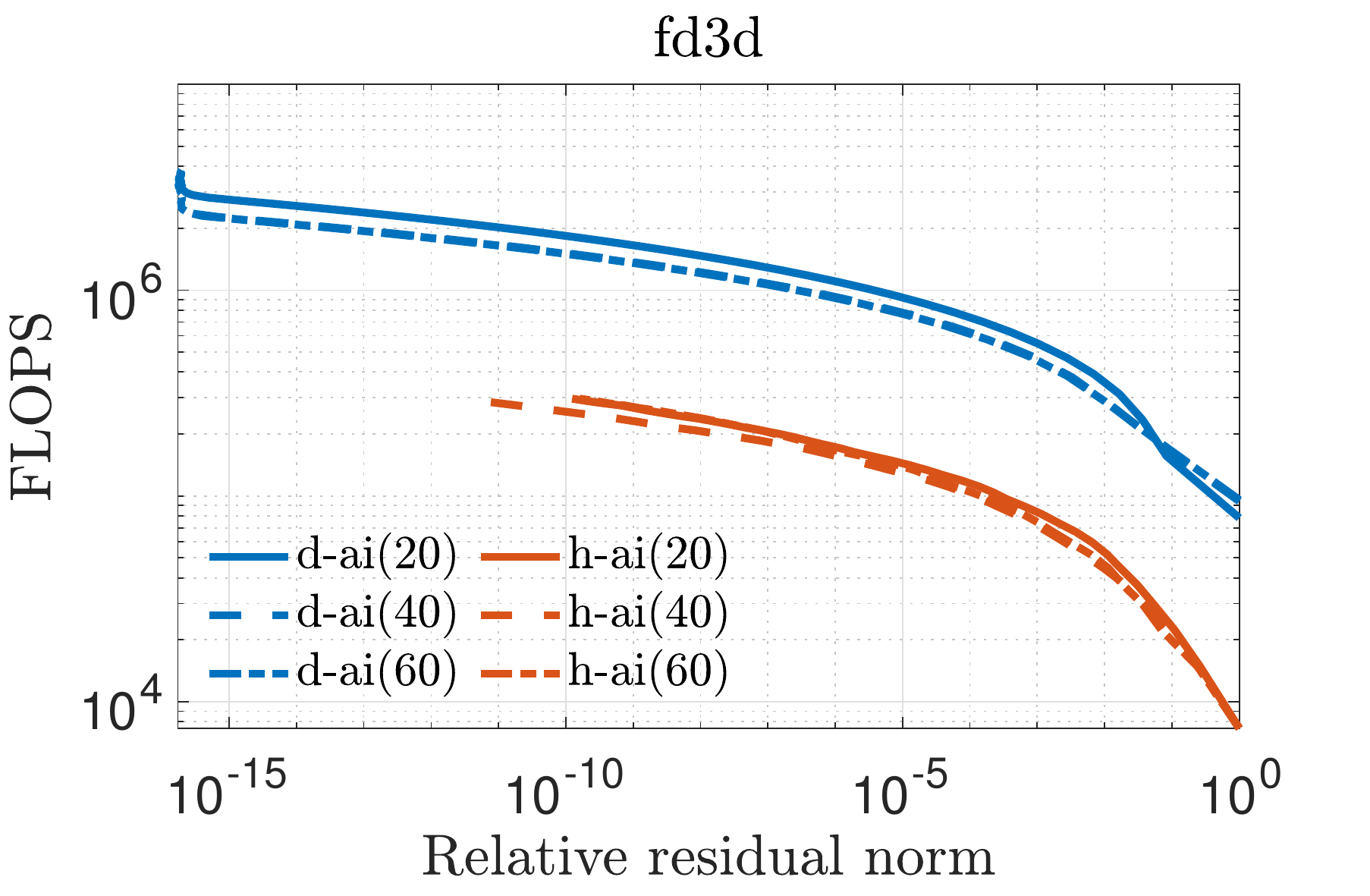}
    \caption{\textit{FLOPs required to reach a certain accuracy by standard and preconditioned 
    Richardson iterations as the maximum number of nonzero entries in $M$ varies according to $\mathtt{nnz}_{\mathrm{AI}}=20\times \mathtt{nnz}(A),\ 40\times \mathtt{nnz}(A)$, and $\mathtt{nnz}_{\mathrm{AI}}=60\times \mathtt{nnz}(A)$.}}\label{fig4}
\vspace*{-0.2in}
\end{figure*}

In our experiments, we mainly compare the following iterative solvers: ($a$) Richardson iterations without preconditioning; ($b$) Richardson iterations with approximate inverse preconditioning applied in the IEEE 754 standard (denoted by ``Richardson+d-ai''); and ($c$) Richardson iterations with approximate inverse preconditioning applied through simulated analog hardware (denoted by 
``Richardson+h-ai''). 
We observe 
from the results in Table~\ref{tab2}
that standard Richardson iterations fail to converge within $m_{\mathrm{it}}=50$ iterations for all three test problems, but converge rapidly with approximate inverse preconditioning.
This is not surprising since the spectral 
radius of the matrix $I-A$, i.e.,  $\rho(I-A)$, is very close to one in all cases. 
The convergence behavior of the two preconditioned variants is almost 
identical for the FE matrices, but hybrid preconditioned Richardson iteration requires more than twice the number iterations compared to the digital variant for the 
FD matrix. Overall, the hybrid preconditioned Richardson 
requires $5\times$ to $10\times$ fewer digital FLOPs compared to the 
purely digital version.

Figure \ref{fig1} plots the relative residual norm after each Richardson iteration both without preconditioning and with the various preconditioners considered
in this paper.
As expected, hybrid 
preconditioning requires more iterations; however, this 
trade-off is highly beneficial because analog hardware allows for a rapid application of the 
preconditioner. This is evident in Figure \ref{fig2} which shows that, on an average, hybrid preconditioning requires substantially fewer digital FLOPs to converge.
For reference, we have also included Richardson iterations with an ILU(0) preconditioner~\cite{saad2003iterative} applied in the IEEE 754 standard. 

Figure \ref{fig4} plots the number of digital FLOPs versus the relative 
residual norm for $\mathtt{nnz}_{\mathrm{AI}}$ values of
$20\times \mathtt{nnz}(A)$, $40\times \mathtt{nnz}(A)$, and $60\times \mathtt{nnz}(A)$. A denser preconditioner speeds up the convergence rate at the cost of additional FLOPs per iteration in the digital implementation. For the square FE mesh, the hybrid implementation yields a greater improvement in FLOPs compared to its digital counterpart as the preconditioner is made denser because the time to apply $M$ on the analog device does not increase with nnz($M$). However, this trend is not observed for the other matrices and diminishes with increasing preconditioner density because the analog noises limit the accuracy of the MVM operation $Mr$ even if $M$ is highly accurate. As a result, the number of iterations does not always decline as nnz($M$) increases.

Figures \ref{fig1}--\ref{fig4} show that Richardson iterations preconditioned with approximate inverses on the hybrid architecture outperform preconditioned Richardson iterations on the digital architecture by greater margins as the condition number (listed in Table~\ref{tab2}) of $A$ grows larger. Figure~\ref{fig4} also shows that our method can use denser preconditioners more effectively for matrices with higher condition numbers. These trends bode well for the potential use of our proposed preconditioning framework on a hybrid architecture in real-life problems that are likely to be larger and more ill-conditioned.

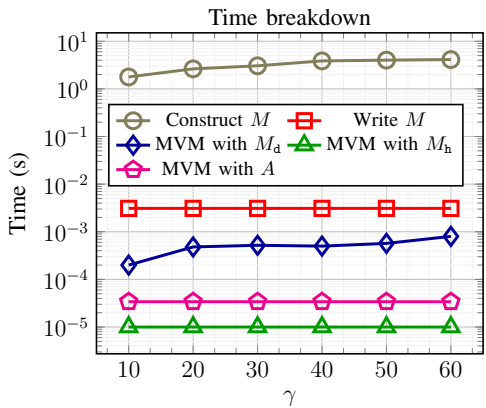
\begin{figure}[htbp]
\centering
\pgfplotsset{every tick label/.append style={font=\large}}
\begin{tikzpicture}[scale=0.75]
\begin{semilogyaxis}[thick,xlabel={\large $\gamma$},	ylabel={\large Time (s)},minor tick num=2,grid=both,
    grid style={line width=.1pt, draw=gray!10},
    major grid style={line width=.2pt,draw=gray!50},legend columns=2,title={\large Time breakdown},
    legend style={at={(0.03,0.65)},anchor=west},
    title style={at={(0.5,1.06)},anchor=north,yshift=-0.1},]
\addplot[black!60!yellow,mark=o,mark options=solid,mark size=4pt,line width=1.5pt] coordinates {
	(10,1.78)    (20,2.63)   (30,3.05)
	(40,3.86)  (50,4.01)  (60,4.12)
};

\addplot[red,mark=square,mark options=solid,mark size=3.6pt,line width=1.5pt] coordinates{
	(10,0.0031)    (20,0.0031)   (30,0.0031)
	(40,0.0031)  (50,0.0031)  (60,0.0031)
};

\addplot[black!40!blue,mark=diamond,mark options=solid,mark size=4.6pt,line width=1.5pt] coordinates{
	(10,2e-4)    (20,4.8e-4)   (30,5.2e-4)
	(40,5e-4)  (50,5.7e-4)  (60,8e-4)
};

\addplot[black!40!green,mark=triangle,mark options=solid,mark size=4.8pt,line width=1.5pt] coordinates{
	(10,1e-5)    (20,1e-5)   (30,1e-5)
	(40,1e-5)  (50,1e-5)  (60,1e-5)
};

\addplot[magenta,mark=pentagon,mark options=solid,mark size=4.3pt,line width=1.5pt] coordinates{
	(10,3.4e-5)    (20,3.4e-5)   (30,3.4e-5)
	(40,3.4e-5)  (50,3.4e-5)  (60,3.4e-5)
};
\legend{Construct $M$, Write $M$, MVM with $M_{\mathtt{d}}$, MVM with $M_{\mathtt{h}}$, MVM with $A$}
\end{semilogyaxis}
\end{tikzpicture}
\caption{\textit{Timings of
parts of Algorithm \ref{alg:richardson} as
the 
maximum number of allowed nonzero entries $\mathtt{nnz}_{\mathrm{AI}}$ in matrix $M$
is
set to $\gamma\times \mathtt{nnz}(A)$}.} \label{breakdown}
\vspace*{-0.1in}
\end{figure}

Figure \ref{breakdown} plots the impact of varying the fill-in factor for the FE square matrix on the time required by the key operations of Algorithm~\ref{alg:richardson}, such as constructing $M$, writing it to the analog crossbar, performing an MVM with it in both the digital and hybrid settings, and performing a digital 
MVM with the coefficient matrix $A$. The sequential time for
the construction of the preconditioner
in MATLAB is reported for academic purposes only; in practice, this step would utilize compiler optimizations and the ample availability of  parallelism~\cite{chow2001parallel,grote1997parallel} to finish much faster. The time for writing $M$ is roughly equivalent to that of performing seven
digital MVMs with it, implying that the benefits of the hybrid architecture start accruing after
seven iterations.
Note that the size of this matrix $M$ is only $625\times 625$.
The current state of analog hardware technology is capable of realizing crossbar arrays of sizes up to $4000\times4000$~\cite{Gokmen2016}. These can accommodate much larger preconditoner matrices for which the analog MVM times would still be similar to those in Figure~\ref{breakdown} and the write times will increase as $O(n)$, but the digital MVM times would be much higher, growing as $O(n^2)$.

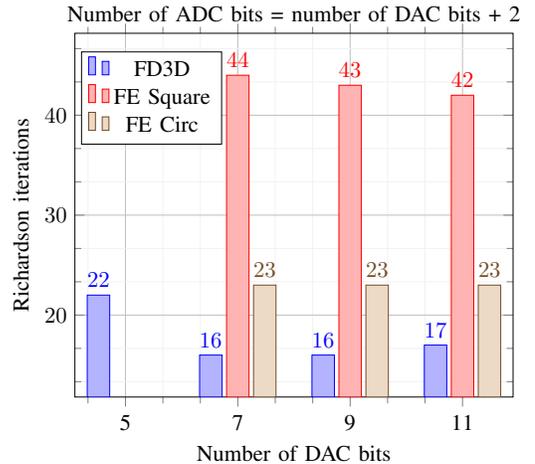
\begin{figure}[htbp]
\centering
\begin{tikzpicture}[scale=0.85]
\begin{axis}[minor tick num=3,
    ybar,
    enlargelimits=0.15,
    ylabel={Richardson iterations},
    xlabel={Number of DAC bits},
    symbolic x coords={5, 7, 9, 11},
    legend style={at={(0.175,0.95)},
    anchor=north,legend columns=-1},
    nodes near coords,
    every node near coord/.append style={font=\tiny},
   nodes near coords align={vertical},
   title={Number of ADC bits = number of DAC bits + 2},
   title style={at={(0.5,1.06)},anchor=north,yshift=-0.1},
   legend columns=1,every node near coord/.append style={font=\normalsize},minor tick num=2,grid=both,
    grid style={line width=.1pt, draw=gray!10},
    major grid style={line width=.2pt,draw=gray!50},
    ]
\addplot coordinates {(5, 22) (7, 16) (9, 16) (11, 17)};
\addplot coordinates {(7, 44) (9, 43) (11, 42)};
\addplot coordinates {(7, 23) (9, 23) (11, 23)};
\legend{FD3D, FE Square, FE Circ}
\end{axis}
\end{tikzpicture}
\caption{\textit{Number of iterations required by hybrid preconditioned Richardson as the 
number of DAC/ADC bits vary. For the 5-bit DAC, no convergence was obtained after 
100 iterations for the FE matrices}.}\label{dacbits}
\vspace*{-0.1in}
\end{figure}

Some hardware parameters,
such as
the number of DAC and ADC bits, can have a profound impact on the time and energy consumption of the analog device, and
therefore
on its overall performance. We vary
the number of DAC and ADC bits and plot 
the results in Figure~\ref{dacbits} 
to justify our choice of 7 DAC bits and 9 ADC bits.
Fewer bits tend to lead to divergence, while
more bits result in increased time and energy consumption with little or no improvement in convergence, 
as the accuracy of the analog computations is limited by other sources of noise.  

\section{Concluding remarks}
\label{sec-conclusion}

With the slowing of Moore's law \cite{mooreslaw2020} for digital microprocessors, there has been considerable interest in exploring alternatives, such as analog computing, for speeding up computationally expensive kernels. While simple analog crossbar arrays have been shown to be effective in many applications involving dense matrices, it has been challenging to address sparse matrix problems, such as solving sparse linear systems, because they do not naturally map to dense crossbar arrays and generally have a low tolerance for the stochastic errors pervasive in analog computing.
This paper proposes, analyzes, and experimentally evaluates a preconditioning framework that exploits inexpensive MVM on analog arrays to substantially reduce the time and energy required for solving an important practical class of sparse linear systems. These and similar efforts are critical for meeting the continually growing computational demands of various applications of sparse solvers and for preparing these solvers for the fast but energy-constrained embedded and exascale~\cite{Exascale2020} systems of the future.

\section*{Acknowledgments}
We would like to thank Tayfun Gokmen, Malte Rasch, and Shashanka Ubaru for helpful discussions and input that substantially contributed to the content of this paper.

\clearpage
\bibliographystyle{IEEEtran}
\bibliography{references}

\begin{thebibliography}{10}
\providecommand{\url}[1]{#1}
\csname url@samestyle\endcsname
\providecommand{\newblock}{\relax}
\providecommand{\bibinfo}[2]{#2}
\providecommand{\BIBentrySTDinterwordspacing}{\spaceskip=0pt\relax}
\providecommand{\BIBentryALTinterwordstretchfactor}{4}
\providecommand{\BIBentryALTinterwordspacing}{\spaceskip=\fontdimen2\font plus
\BIBentryALTinterwordstretchfactor\fontdimen3\font minus
  \fontdimen4\font\relax}
\providecommand{\BIBforeignlanguage}[2]{{%
\expandafter\ifx\csname l@#1\endcsname\relax
\typeout{** WARNING: IEEEtran.bst: No hyphenation pattern has been}%
\typeout{** loaded for the language `#1'. Using the pattern for}%
\typeout{** the default language instead.}%
\else
\language=\csname l@#1\endcsname
\fi
#2}}
\providecommand{\BIBdecl}{\relax}
\BIBdecl

\bibitem{saad2003iterative}
Y.~Saad, \emph{Iterative methods for sparse linear systems}.\hskip 1em plus
  0.5em minus 0.4em\relax SIAM, 2003.

\bibitem{bertaccini2016sparse}
D.~Bertaccini and S.~Filippone, ``Sparse approximate inverse preconditioners on
  high performance {GPU} platforms,'' \emph{Computers \& Mathematics with
  Applications}, vol.~71, no.~3, pp. 693--711, 2016.

\bibitem{abdelfattah2020survey}
A.~Abdelfattah, H.~Anzt, E.~G. Boman, E.~Carson, T.~Cojean, J.~Dongarra,
  M.~Gates, T.~Gr{\"u}tzmacher, N.~J. Higham, S.~Li \emph{et~al.}, ``A survey
  of numerical methods utilizing mixed precision arithmetic,'' \emph{arXiv
  preprint arXiv:2007.06674}, 2020.

\bibitem{oo2020accelerating}
K.~L. Oo and A.~Vogel, ``Accelerating geometric multigrid preconditioning with
  half-precision arithmetic on {GPU}s,'' \emph{arXiv preprint
  arXiv:2007.07539}, 2020.

\bibitem{baboulin2009accelerating}
M.~Baboulin, A.~Buttari, J.~Dongarra, J.~Kurzak, J.~Langou, J.~Langou,
  P.~Luszczek, and S.~Tomov, ``Accelerating scientific computations with mixed
  precision algorithms,'' \emph{Computer Physics Communications}, vol. 180,
  no.~12, pp. 2526--2533, 2009.

\bibitem{haidar2018harnessing}
A.~Haidar, S.~Tomov, J.~Dongarra, and N.~J. Higham, ``Harnessing {GPU} tensor
  cores for fast fp16 arithmetic to speed up mixed-precision iterative
  refinement solvers,'' in \emph{SC18: Int. Conf. for High Perf. Computing,
  Networking, Storage and Analysis}.\hskip 1em plus 0.5em minus 0.4em\relax
  IEEE, 2018, pp. 603--613.

\bibitem{anzt2017preconditioned}
H.~Anzt, M.~Gates, J.~Dongarra, M.~Kreutzer, G.~Wellein, and M.~K{\"o}hler,
  ``Preconditioned krylov solvers on {GPU}s,'' \emph{Parallel Computing},
  vol.~68, pp. 32--44, 2017.

\bibitem{haidar2020mixed}
A.~Haidar, H.~Bayraktar, S.~Tomov, J.~Dongarra, and N.~J. Higham,
  ``Mixed-precision iterative refinement using tensor cores on {GPU}s to
  accelerate solution of linear systems,'' \emph{Proceedings of the Royal
  Society A}, vol. 476, no. 2243, p. 20200110, 2020.

\bibitem{hu2016dot}
M.~Hu \emph{et~al.}, ``Dot-product engine for neuromorphic computing:
  Programming 1t1m crossbar to accelerate matrix-vector multiplication,'' in
  \emph{2016 53nd ACM/EDAC/IEEE Design Automation Conference (DAC)}.\hskip 1em
  plus 0.5em minus 0.4em\relax IEEE, 2016, pp. 1--6.

\bibitem{xia2016technological}
L.~Xia, P.~Gu, B.~Li, T.~Tang, X.~Yin, W.~Huangfu, S.~Yu, Y.~Cao, Y.~Wang, and
  H.~Yang, ``Technological exploration of rram crossbar array for matrix-vector
  multiplication,'' \emph{Journal of Computer Science and Technology}, vol.~31,
  no.~1, pp. 3--19, 2016.

\bibitem{Sebastian-NN}
A.~Sebastian, M.~L.~G. Le~Gallo, R.~Khaddam-Aljameh, and E.~Eleftheriou,
  ``Memory devices and applications for in-memory computing,'' \emph{Nature
  Nanotechnology}, vol.~15, pp. 529--544, 2020.

\bibitem{Haensch2019}
W.~Haensch, T.~Gokmen, and R.~Puri, ``The next generation of deep learning
  hardware: Analog computing,'' \emph{Proceedings of the IEEE}, vol. 107, pp.
  108--122, 2019.

\bibitem{RAPA2019}
M.~J. Rasch, T.~Gokmen, M.~Rigotti, and W.~Haensch, ``{RAPA-ConvNets}: Modified
  convolutional networks for accelerated training on architectures with analog
  arrays,'' \emph{Frontiers in Neuroscience}, vol.~13, p. 753, 2019.

\bibitem{Ambrogio2018}
S.~Ambrogio \emph{et~al.}, ``Equivalent-accuracy accelerated neural-network
  training using analogue memory,'' \emph{Nature}, vol. 558, pp. 60--67, 2018.

\bibitem{Fumorola2016}
A.~Fumarola, P.~Narayanan, L.~L. Sanches, S.~Sidler, J.~Jang, and K.~Moon,
  ``Accelerating machine learning with non-volatile memory: exploring device
  and circuit tradeoffs,'' in \emph{IEEE International Conference on Rebooting
  Computing (ICRC)}, 2016, pp. 1--8.

\bibitem{Gokmen2016}
T.~Gokmen and Y.~Vlasov, ``Acceleration of deep neural network training with
  resistive cross-point devices: Design considerations,'' \emph{Frontiers in
  Neuroscience}, vol.~10, 2016.

\bibitem{Burr2016}
G.~W. Burr, R.~M. Shelby, A.~Sebastian, S.~Kim, and S.~Sidler, ``Neuromorphic
  computing using non-volatile memory,'' \emph{Advances in Physics}, vol.~2,
  pp. 89--124, 2016.

\bibitem{Bojnordi2016}
M.~N. Bojnordi and E.~Ipek, ``Memristive {B}oltzmann machine: A hardware
  accelerator for combined optimization and deep learning,'' in \emph{Int.
  Symp. on High Performance Computer Architecture (HPCA)}, 2016.

\bibitem{Shafice2016}
A.~Shafice \emph{et~al.}, ``{ISAAC}: A convolutional neural network accelerator
  with in-situ analog arithmetic in crossbars,'' in \emph{International
  Symposium on Computer Architecture (ISCA)}, 2016.

\bibitem{Feinberg2021}
B.~Feinberg, R.~Wong, T.~P. Xiao, C.~H. Bennett, J.~N. Rohan, E.~G. Boman,
  M.~J. Marinella, S.~Agarwal, and E.~Ipek, ``An analog preconditioner for
  solving linear systems,'' in \emph{2021 IEEE Int. Symp. on High-Performance
  Computer Architecture (HPCA)}.\hskip 1em plus 0.5em minus 0.4em\relax IEEE,
  pp. 761--774.

\bibitem{Feinberg2018}
B.~Feinberg, U.~K.~R. Vengalam, N.~Whitehair, S.~Wang, and E.~Ipek, ``Enabling
  scientific computing on memristive accelerators,'' in \emph{International
  Symposium on Computer Architecture (ISCA)}, 2018, pp. 367--382.

\bibitem{zidan2018general}
M.~A. Zidan, Y.~Jeong, J.~Lee, B.~Chen, S.~Huang, M.~J. Kushner, and W.~D. Lu,
  ``A general memristor-based partial differential equation solver,''
  \emph{Nature Electronics}, vol.~1, no.~7, pp. 411--420, 2018.

\bibitem{sun2019solving}
Z.~Sun, G.~Pedretti, E.~Ambrosi, A.~Bricalli, W.~Wang, and D.~Ielmini,
  ``Solving matrix equations in one step with cross-point resistive arrays,''
  \emph{Proceedings of the National Academy of Sciences}, vol. 116, no.~10, pp.
  4123--4128, 2019.

\bibitem{richter2015memristive}
I.~Richter, K.~Pas, X.~Guo, R.~Patel, J.~Liu, E.~Ipek, and E.~G. Friedman,
  ``Memristive accelerator for extreme scale linear solvers,'' in
  \emph{Government Microcircuit Applications \& Critical Technology Conference
  (GOMACTech)}, 2015.

\bibitem{LeGallo2018}
M.~Le~Gallo, A.~Sebastian, R.~Mathis, M.~Manica, H.~Giefers, T.~Tuma, C.~Bekas,
  A.~Curioni, and E.~Eleftheriou, ``Mixed-precision in-memory computing,''
  \emph{Nature Electronics}, vol.~1, no.~4, pp. 246--253, 2018.

\bibitem{zhu2021fixed}
Z.~Zhu, A.~B. Klein, G.~Li, and S.~Pang, ``Fixed-point iterative linear inverse
  solver with extended precision,'' \emph{arXiv preprint arXiv:2105.02106},
  2021.

\bibitem{Exascale2020}
H.~Antz \emph{et~al.}, ``Preparing sparse solvers for exascale computing,''
  \emph{Philosophical Transactions of the Royal Society of London A:
  Mathematical, Physical and Engineering Sciences}, vol. 378, no. 20190053,
  2020.

\bibitem{Composable2012}
J.~Brown, M.~G. Knepley, D.~A. May, L.~C. McInnes, and B.~Smith, ``Composable
  linear solvers for multiphysics,'' in \emph{11th International Symposium on
  Parallel and Distributed Computing}, 2012, pp. 55--62.

\bibitem{Parco2014}
L.~C. McInnes, B.~Smith, H.~Zhang, and R.~T. Mills, ``Hierarchical and nested
  {Krylov} methods for extreme-scale computing,'' \emph{Parallel Computing},
  vol.~40, no.~1, pp. 17--31, 2014.

\bibitem{chua:1971}
L.~Chua, ``Memristor-the missing circuit element,'' \emph{IEEE Transactions on
  Circuit Theory}, vol.~18, no.~5, pp. 507--519, 1971.

\bibitem{benzi1999bounds}
M.~Benzi and G.~H. Golub, ``Bounds for the entries of matrix functions with
  applications to preconditioning,'' \emph{BIT Numerical Mathematics}, vol.~39,
  no.~3, pp. 417--438, 1999.

\bibitem{demko1984decay}
S.~Demko, W.~F. Moss, and P.~W. Smith, ``Decay rates for inverses of band
  matrices,'' \emph{Mathematics of computation}, vol.~43, no. 168, pp.
  491--499, 1984.

\bibitem{benzi1999comparative}
M.~Benzi and M.~Tuma, ``A comparative study of sparse approximate inverse
  preconditioners,'' \emph{Applied Numerical Mathematics}, vol.~30, no. 2-3,
  pp. 305--340, 1999.

\bibitem{chow2001parallel}
E.~Chow, ``Parallel implementation and practical use of sparse approximate
  inverse preconditioners with a priori sparsity patterns,'' \emph{The
  International Journal of High Performance Computing Applications}, vol.~15,
  no.~1, pp. 56--74, 2001.

\bibitem{grote1997parallel}
M.~J. Grote and T.~Huckle, ``Parallel preconditioning with sparse approximate
  inverses,'' \emph{SIAM Journal on Scientific Computing}, vol.~18, no.~3, pp.
  838--853, 1997.

\bibitem{saad1986gmres}
Y.~Saad and M.~H. Schultz, ``{GMRES}: A generalized minimal residual algorithm
  for solving nonsymmetric linear systems,'' \emph{SIAM J. on scientific and
  statistical computing}, vol.~7, no.~3, pp. 856--869, 1986.

\bibitem{saad1993flexible}
Y.~Saad, ``A flexible inner-outer preconditioned {GMRES} algorithm,''
  \emph{SIAM J. on Scientific Computing}, vol.~14, no.~2, pp. 461--469, 1993.

\bibitem{richardson1911ix}
L.~F. Richardson, ``{IX.} {T}he approximate arithmetical solution by finite
  differences of physical problems involving differential equations, with an
  application to the stresses in a masonry dam,'' \emph{Philosophical
  Transactions of the Royal Society of London A: Mathematical, Physical and
  Engineering Sciences}, vol. 210, no. 459-470, pp. 307--357, 1911.

\bibitem{Lawson:1979:BLA}
C.~L. Lawson, R.~J. Hanson, D.~R. Kincaid, and F.~T. Krogh, ``{Basic Linear
  Algebra Subprograms} for {Fortran} usage,'' \emph{{ACM} Transactions on
  Mathematical Software}, vol.~5, no.~3, pp. 308--323, 1979.

\bibitem{Moler}
C.~Moler, ``Iterative refinement in floating point,'' \emph{Annals of the ACM},
  vol.~14, no.~2, pp. 316--321, 1967.

\bibitem{rasch2021flexible}
M.~J. Rasch, D.~Moreda, T.~Gokmen, M.~L. Gallo, F.~Carta, C.~Goldberg, K.~E.
  Maghraoui, A.~Sebastian, and V.~Narayanan, ``A flexible and fast pytorch
  toolkit for simulating training and inference on analog crossbar arrays,''
  2021.

\bibitem{mooreslaw2020}
J.~Shalf, ``The future of computing beyond {M}oore’s law,'' \emph{Phil.
  Trans. R. Soc. A.}, vol. 378, no. 20190061, 2020.

\bibitem{Wilkinson}
J.~Wilkinson, \emph{Rounding Errors in Algebraic Processes}.\hskip 1em plus
  0.5em minus 0.4em\relax Prentice Hall, 1963.

\bibitem{Merikoski}
J.~Merikoski and R.~Kumar, ``Lower bounds for the spectral norm,''
  \emph{Journal of Inequalities in Pure and Applied Mathematics}, vol.~6,
  no.~4, 2005, article 124.

\bibitem{Billingsley}
P.~Billingsley, \emph{Probability and Measure}, 3rd~ed.\hskip 1em plus 0.5em
  minus 0.4em\relax John Wiley \& Sons, 1995.

\end{thebibliography}

\onecolumn
\appendices

\section{Error and Convergence Analysis}
\label{app:IR}

\subsection{Error in MVM Due to Device Noise}
\label{app:DeviceNoise}

Performing the MVM operation $y = M r$ with a matrix $M \in \mathbb{R}^{n\times n}$ and vector $r \in \mathbb{R}$ on an analog crossbar array involves multiple sources of nondeterministic noise so that the output $\widehat{y} \in \mathbb{R}$ is equivalent to a low precision approximation of $y$. Writing $M$ to the crossbar array incurs multiplicative and additive write noises $N^{Wm} \in \mathbb{R}^{n\times n}$ and $N^{Wa} \in \mathbb{R}^{n\times n}$, respectively, and the actual conductance values at the crosspoints of the array are given by $\widehat{M} = M\odot(I + N^{Wm}) + N^{Wa}$, where $\odot$ denotes element-wise multiplication. Similarly, digital-to-analog conversion (DAC) of the vector $r$ into voltage pulses suffers from multiplicative and additive input noises $N^{Im} \in \mathbb{R}^n$ and $N^{Ia} \in \mathbb{R}^n$, respectively. As a result, the matrix $\widehat{M}$ is effectively multiplied by a perturbed version of $r$ given by $\widehat{r} = r\odot(\mathbf{1} + N^{Im}) + N^{Ia}$, where $\mathbf{1}$ is a vector of all ones.

A characteristic equation to describe the output $\widehat{y}$ of an analog MVM $Mr$ 
is given by
\begin{equation}
\label{eqn-error-app1}
\widehat{y} = \left((M\odot(I + N^{Wm}) + N^{Wa})(r\odot(\mathbf{1} + N^{Im}) + N^{Ia})\right)\odot (\mathbf{1} + N^{Om}) + N^{Oa},
\end{equation}
where $N^{Om} \in \mathbb{R}^n$ and $N^{Oa} \in \mathbb{R}^n$ denote the multiplicative and additive components of the output noise, respectively, which reflects the inherent inexactness of the multiplication based on circuit laws and current integration, as well as the loss of precision in the ADC conversion of the result vector. 

Equation~\eqref{eqn-error-app1} can be rearranged as
\[
\widehat{y} = \left((M\odot(I+N^{Wm})+N^{Wa})r\odot(\mathbf{1}+N^{Im})\right)\odot(\mathbf{1}+N^{Om}) + (M\odot(I + N^{Wm})N^{Ia} + N^{Wa}N^{Ia})\odot(\mathbf{1}+N^{Om}) + N^{Oa} ,
\]
or more concisely
\begin{equation}
\label{eqn-error-app2}
\widehat{y} = \left(M\odot(I+N^{Wm}) + N^{Wa}\right) r\odot(\mathbf{1}+N^{Im})\odot(\mathbf{1}+N^{Om}) + \widehat{N}^{Oa},
\end{equation}
where $\widehat{N}^{Oa}$ is the overall effective additive noise $(M\odot(I + N^{Wm})N^{Ia} + N^{Wa}N^{Ia})\odot(\mathbf{1}+N^{Om}) + N^{Oa}$ and captures the lower order noise terms. 
The $i$th entry $\widehat{y}[i]$ of $\widehat{y}$ in Equation~\eqref{eqn-error-app2} is given by
\[
\widehat{y}[i] = \left(\sum_{j=1}^n\left(M[i,j](1+N^{Wm}[i,j]) + N^{Wa}[i,j]\right)r[j](1+N^{Im}[j])(1+N^{Om}[i])\right) + \widehat{N}^{Oa}[i] ,
\]
and upon ignoring the products of components of $N^{Wm}$, $N^{Im}$ and $N^{Om}$, we then have
\begin{equation}
\label{eqn-error-app3}
\widehat{y}[i] = \left(\sum_{j=1}^n\left(M[i,j](1 + N^{Wm}[i,j] + N^{Im}[j] + N^{Om}[i]) + N^{Wa}[i,j]\right)r[j]\right) + \widehat{N}^{Oa}[i].
\end{equation}
Rewriting Equation~\eqref{eqn-error-app3} in matrix form yields
\[
\widehat{y} = \left(M\odot(I + N^{Wm}+\mathbf{1}\otimes N^{Im} + N^{Om}\otimes\mathbf{1}) + N^{Wa}\right)r + \widehat{N}^{Oa},
\]
where $\otimes$ denotes outer product, or more concisely
\begin{equation}
\label{eqn-error-app4}
\widehat{y} = (M + E)r + \widehat{N}^{Oa},
\end{equation}
where the nondeterministic error matrix $E$ is given by
\begin{equation}
\label{eqn-error-app5}
E = M\odot(N^{Wm}+\mathbf{1}\otimes N^{Im} + N^{Om}\otimes\mathbf{1}) + N^{Wa}.
\end{equation}

\subsection{Wilkinson and Moler's Deterministic Analysis of Iterative Refinement from a Digital Perspective}
\label{app:IR-digital}
The iterative refinement (IR) scheme, proposed by Wilkinson in 1948~\cite{Wilkinson}, is often used to reduce errors in solving linear systems $Ax=b$, especially when the matrix $A$ is ill-conditioned.
This IR approach consists of the use of an inaccurate solver for the most expensive operations, namely solving the residual equation $Ad = r \triangleq b-Ax$ (with time complexity $O(n^3)$),
and then updating the estimate $x$ and computing the next residual (both with time complexity $O(n^2)$) using forms of higher precision arithmetic.
Error bounds for the IR algorithm were subsequently analyzed by Moler in 1967~\cite{Moler}.
In particular, it was shown by Wilkinson and Moler that if the condition number of the matrix $A$ is not too large and the residual is computed at double the working precision, then the IR algorithm will converge to the true solution
within working precision.

More precisely, the IR algorithm comprises the following steps:
\begin{enumerate}
\item Solve $Ax = b$;
\item Compute residual $r_m = b-Ax_m$;
\item Solve $Ad_m = r_m$ using a basic (inaccurate computation) method;
\item Update $x_{m+1} = x_m + d_m$;
\item Repeat until stopping criteria satisfied.
\end{enumerate}
Although there are many variants of the above IR algorithm, our primary focus here is on the use of analog hardware to compute the solution of the linear system $Ad_m = r_m$ in step $3$,
rendering the benefits of greater efficiency and/or lower power consumption, but at the expense of lower and noisier precision.
All other steps in the above algorithm are assumed to be computed on a digital computer at some form of higher precision.

We end this brief
summary of
previous work
by providing a high-level overview of the 
original
error and convergence analysis
due to
Moler~\cite{Moler}.
Consider the IR algorithm and focus on the following three steps for each iteration $m$:
\begin{enumerate}
\item $r_m = b - Ax_m + c_m$;
\item $A(I+F_m)d_m = r_m$;
\item $x_{m+1} = x_m + d_m + g_m$.
\end{enumerate}
Here, $c_m$, $F_m$ and $g_m$ are corrections that allow for the above equations to hold exactly, since the actual computations are performed on hardware that is inexact. For now, to simplify the exposition of our analysis, let us assume $g_m = c_m = 0$, i.e., steps $1$ and $3$ are computed with perfect precision; positive instances of these precision constants can be readily incorporated subsequently in our analysis.

Suppose it can be shown that $\|F_m\| < 1, \; \forall m$, and hence the matrices $I+F_m$ are nonsingular. Furthermore, we have $$(I+F_m)^{-1}F_m = \sum_{i=0}^\infty (-1)^i F_m^{i+1},$$
which implies
\begin{equation}\label{eq:Fm-upperbound}
    \|(I+F_m)^{-1}F_m\| \leq \sum_{i=0}^\infty  \|F_m\|^{i+1} = \dfrac{\|F_m\|}{1-\|F_m\|}.
\end{equation}
Then, letting $x = A^{-1}b$ be the true solution, it is possible to algebraically establish that
\[ x_{m+1} - x = (I+F_m)^{-1}F_m(x_m-x) ,\]
from which it follows
\[ \|x_{m+1} - x \| \leq \dfrac{\|F_m\|}{1-\|F_m\|}\|x_m-x\| .\]
This therefore leads to a contraction when $\|F_m\|$ is sufficiently small
(in particular, $\|F_m\| < 1/2, \; \forall m$),
thus guaranteeing convergence with a rate corresponding to the above equation.

\subsection{Analysis of Iterative Refinement from an Analog Perspective}
\label{app:IR-analog}
Adapting the above error and convergence analysis due to Moler~\cite{Moler} for our objectives herein,
we start by modeling the writing of the $n\times n$ matrix $A$ into the analog hardware before beginning the
$3$-step iterative process.
This results in a one-time write error that we model as
\begin{equation}\label{eq:widehatA}
\widetilde{A} = A + \widetilde{J} ,
\end{equation}
where $\widetilde{J}$ is a matrix whose elements are independent and identically distributed (i.i.d.) according to
a general distribution $\cD_J(\mu_J,\sigma_J^2)$ (not necessarily Gaussian) having finite mean $\mu_J$ and
finite variance $\sigma_J^2$.
For the analog devices under consideration~\cite{rasch2021flexible}, $\mu_J = 0$.

Next, we consider step $2$ of the $3$-step iterative process in Appendix~\ref{app:IR-digital} as being the only one performed on inaccurate (analog) hardware.
We want to model in a general fashion the overall noise that is incurred when solving the linear system in step $2$ of each iteration $m$.
To this end, let us assume that the computation of $d_m$ in step $2$ of iteration $m$ 
satisfies
\begin{equation}\label{eq:step2}
(\widetilde{A}+\widetilde{K}_m)d_m = r_m,
\end{equation}
where $\widetilde{K}_m$ are matrices whose elements are i.i.d.\ according to a general distribution $\cD_K(\mu_K,\sigma_K^2)$
(not necessarily Gaussian) having finite mean $\mu_K$ and finite variance $\sigma_K^2$.

We seek to establish bounds on $\|\widetilde{K}_m\|$.
Even though all matrix norms are equivalent in a topological sense, it is convenient and prudent for our purposes to choose matrix norms with which
it is easy to work.
Let us consider the $\max$ norm $\|\cdot \|_{\max}$, the Euclidean distance operator norm $\|\cdot \|_2$ and the Frobenius
matrix norm $\|\cdot \|_F$, taking advantage of the known bounds
$$\sqrt{\dfrac{|r_1|^2 + \cdots + |r_n|^2}{n}} \leq \|\widetilde{K}_m\|_{\max} \leq \|K_m\|_2 \leq \|\widetilde{K}_m\|_F ,$$
where $r_1, \ldots, r_n$ are the row sums of $\widetilde{K}_m$;
refer to~\cite{Merikoski} for this lower bound.
Focusing on the norm $\|\cdot\|$ being either $\|\cdot\|_{\max}$ or $\|\cdot\|_2$, we then derive
\begin{align}
\ex(\|\widetilde{K}_m\|) \leq \ex(\|\widetilde{K}_m\|_F) &= \ex\left( \sqrt{\sum_{ij} (\widetilde{K}_m)_{ij}^2}\right) \nonumber \\
& \leq \sqrt{\ex(\sum_{ij} (\widetilde{K}_m)_{ij}^2)} = \sqrt{\sum_{ij} \ex((\widetilde{K}_m)_{ij}^2)}
= \sqrt{n^2 \ex((\widetilde{K}_m)_{11}^2)} = \sqrt{n^2 (\mu_K^2 + \sigma_K^2)} \nonumber \\
& = n \sqrt{\mu_K^2 + \sigma_K^2} ,
\label{eq:derivationK}
\end{align}
where
the first inequality follows from the above matrix-norm upper bound,
the first equality is by definition of the Frobenius matrix norm,
the second inequality is due to Jensen's inequality,
the second equality is a basic property of the expectation operator,
the third equality is due to the i.i.d.\ assumption,
the fourth equality is by definition of the second moment,
the $(i,j)$th element of the matrix $\widetilde{K}_m$ is denoted by ${(\widetilde{K}_m)}_{ij}$,
and expectation is with respect to $\cD_K(\cdot,\cdot)$.

Let us next consider the variance of $\|\widetilde{K}_m\|$, more specifically $\var(\|\widetilde{K}_m\|) = \ex(\|\widetilde{K}_m\|^2) - \ex(\|\widetilde{K}_m\|)^2$.
To this end, we focus initially on the second moment $\ex(\|\widetilde{K}_m\|^2)$ and follow along the lines of the derivation of Equation~\eqref{eq:derivationK} above,
to obtain
\begin{align}
\ex(\|\widetilde{K}_m\|^2) \leq \ex(\|\widetilde{K}_m\|^2_F) &= \ex\left( \left(\sqrt{\sum_{ij} (\widetilde{K}_m)_{ij}^2}\right)^2 \right) 
= \ex\left( \sum_{ij} (\widetilde{K}_m)_{ij}^2 \right) = \sum_{ij} \ex( (\widetilde{K}_m)_{ij}^2 ) \nonumber \\
& = n^2 \ex((\widetilde{K}_m)_{11}^2) = n^2 (\mu_K^2 + \sigma_K^2) ,
\label{eq:derivationK:var2}
\end{align}
where expectation is again with respect to $\cD_K(\cdot,\cdot)$.
Now we turn to a lower bound on $\ex(\|\widetilde{K}_m\|)$, for which we exploit the above matrix-norm lower bound to conclude
\begin{align*}
\ex(\|\widetilde{K}_m\|) & \geq \ex\left(\sqrt{\dfrac{ \sum_i |(\sum_j (\widetilde{K}_m)_{ij})|^2 }{n}}\right).
\end{align*}
Then we derive for the numerator on the right hand side
\begin{align*}
\sum_i |(\sum_j (\widetilde{K}_m)_{ij})|^2 & = \sum_i \bigg(n \sum_j \dfrac{(\widetilde{K}_m)_{ij}}{n}\bigg)^2 \\
& = n^2 \left\{ \sum_i \bigg(\sum_j \dfrac{(\widetilde{K}_m)_{ij}}{n} - \sum_k\sum_j \dfrac{(\widetilde{K}_m)_{kj}}{n^2} \bigg)^2 
+ 2\bigg(\sum_i\sum_j \dfrac{(\widetilde{K}_m)_{ij}}{n^2} \bigg) \bigg(\sum_i\sum_j \dfrac{(\widetilde{K}_m)_{ij}}{n}\bigg) 
\right. \\
& \qquad\qquad\qquad\qquad \left. - n\bigg(\sum_i\sum_j \bigg(\dfrac{(\widetilde{K}_m)_{ij}}{n^2} \bigg)\bigg)^2 \right\} \\
& = n^2 \left\{ \sum_i \bigg(\sum_j \dfrac{(\widetilde{K}_m)_{ij}}{n} - \sum_k\sum_j \dfrac{(\widetilde{K}_m)_{kj}}{n^2} \bigg)^2 
+ 2n\bigg(\sum_i\sum_j \dfrac{(\widetilde{K}_m)_{ij}}{n^2} \bigg)^2 - n\bigg(\sum_i\sum_j \bigg(\dfrac{(\widetilde{K}_m)_{ij}}{n^2} \bigg)\bigg)^2 \right\} \\
& = n^2 \left\{ \sum_i \bigg(\sum_j \dfrac{(\widetilde{K}_m)_{ij}}{n} - \sum_k\sum_j \dfrac{(\widetilde{K}_m)_{kj}}{n^2} \bigg)^2 
+ n \bigg( \sum_i\sum_j \dfrac{(\widetilde{K}_m)_{ij}}{n^2} \bigg)^2 \right\} \\
& \geq n^3 \bigg( \sum_i\sum_j \dfrac{(\widetilde{K}_m)_{ij}}{n^2} \bigg)^2 ,
\end{align*}
from which we obtain
\begin{align}
\ex(\|\widetilde{K}_m\|) & \geq \ex\left(\sqrt{\dfrac{ \sum_i |(\sum_j (\widetilde{K}_m)_{ij})|^2 }{n}}\right) \geq \ex\left(\sqrt{ \dfrac{ n^3 \big( \sum_i\sum_j (\widetilde{K}_m)_{ij}/n^2 \big)^2 }{n} }\right) \nonumber \\
& = \ex\left(\sqrt{ n^2 \bigg( \sum_i\sum_j \dfrac{(\widetilde{K}_m)_{ij}}{n^2} \bigg)^2 } \right) \geq n \ex\bigg(\Big|\sum_i\sum_j \dfrac{(\widetilde{K}_m)_{ij}}{n^2} \Big|\bigg) \nonumber \\
& \geq n \ex\bigg(\dfrac{1}{n^2}\sum_i\sum_j (\widetilde{K}_m)_{ij} \bigg) = n \mu_K .
\label{eq:derivationK:varlb}
\end{align}
Upon combining Equations~\eqref{eq:derivationK:var2} and \eqref{eq:derivationK:varlb}, this yields
\begin{equation}
\var(\|\widetilde{K}_m\|) = \ex(\|\widetilde{K}_m\|^2) - \ex(\|\widetilde{K}_m\|)^2
\leq n^2 (\mu_K^2 + \sigma_K^2) - n^2 \mu_K^2 = n^2 \sigma_K^2 .
\label{eq:derivationK:varub}
\end{equation}

Hence, under the assumption that the elements of $\widetilde{K}_m$ are i.i.d.\ random variables with general distribution $\cD_K(\mu_K,\sigma_K^2)$,
it follows from the above analysis that
\begin{equation}
0 \leq \ex(\|\widetilde{K}_m\|) \leq n\sqrt{ \mu_K^2 + \sigma_K^2 } \qquad 
\mbox{ and } \qquad \var(\|\widetilde{K}_m\|) \leq n^2 \sigma_K^2 .
\label{eq:boundsK}
\end{equation}

Next we turn to complete the error and convergence analysis that covers the iterative process with per-iteration errors modeled via $\widetilde{K}_m$.
From step $2$ of the original $3$-step IR algorithm together with Equation~\eqref{eq:step2}, we have
\begin{equation*}
\widetilde{F}_m = \widetilde{A}^{-1} \widetilde{K}_m .
\end{equation*}
Letting $\kappa(\widetilde{A})$ denote an upper bound on $\ex(\| \widetilde{A}^{-1} \|)$, together with Equation~\eqref{eq:boundsK} and recalling that $\widetilde{J}$ and $\widetilde{K}_m$ are independent by definition,
we then derive
\begin{equation}\label{bnd:EF}
\ex(\|\widetilde{F}_m\|) \leq \ex(\|\widetilde{A}^{-1}\|) \, \ex(\|\widetilde{K}_m\| ) \leq \kappa(\widetilde{A}) \left(n\sqrt{ \mu_K^2 + \sigma_K^2} \right)
\end{equation}
and
\begin{align}\label{bnd:VF}
\var(\|\widetilde{F}_m\|) & \leq \var( \|\widetilde{A}^{-1}\| \, \|\widetilde{K}_m\| ) = \|\widetilde{A}^{-1}\|^2 \, \var(\|\widetilde{K}_m\|) \nonumber \\
& \leq \kappa(\widetilde{A})^2 \left( n^2 \sigma_K^2 \right),
\end{align}
which also renders
\begin{align*}
    \ex(\|\widetilde{F}_m\|^2) & \leq \kappa(\widetilde{A})^2 \, n^2 \sigma_K^2 + \kappa(\widetilde{A})^2 n^2 (\mu_K^2 + \sigma_K^2)
= \kappa(\widetilde{A})^2 \, n^2 ( \mu_K^2 + 2 \sigma_K^2) .
\end{align*}

Recall from the original analysis above due to Moler that
\begin{equation}
x_{m+1} - x = (I+F_m)^{-1}F_m(x_m-x) ,
\label{eq:Moler-recall}
 \end{equation}
from which we have by definition for our purposes
$$ \pr\left( \dfrac{\|\widetilde{x}_{m+1} - \widetilde{x} \|}{\|\widetilde{x}_m-\widetilde{x}\|} > z \right) \leq \pr\left( \|(I+\widetilde{F}_m)^{-1}\widetilde{F}_m\| > z \right) , \qquad \forall z , $$
where $\widetilde{x}$ denotes the solution to $\widetilde{A}\widetilde{x}=b$.
We further obtain from Equation~\eqref{eq:Moler-recall}
$$\widetilde{x}_{m+1}-\widetilde{x} =
(I+\widetilde{F}_m)^{-1}\widetilde{F}_m (\widetilde{x}_m-\widetilde{x}) = \left(\prod_{k=0}^m (I+\widetilde{F}_k)^{-1}\widetilde{F}_k \right) (\widetilde{x}_0-\widetilde{x}),$$
which then leads to
\begin{align}
\|\widetilde{x}_{m+1}-\widetilde{x}\| & \leq \left\|\prod_{k=0}^m (I+\widetilde{F}_k)^{-1}\widetilde{F}_k \right\| \|\widetilde{x}_0-\widetilde{x}\|
\leq \prod_{k=0}^m \|(I+\widetilde{F}_k)^{-1}\widetilde{F}_k \| \|\widetilde{x}_0-\widetilde{x}\| \label{eq:diff-norm} \\
\log\|\widetilde{x}_{m+1}-\widetilde{x}\|
&\leq \sum_{k=0}^m \log\|(I+\widetilde{F}_k)^{-1}\widetilde{F}_k \| + \log\|\widetilde{x}_0-\widetilde{x}\| , \label{eq:log-diff-norm}
\end{align}
where the second inequality follows from the triangle inequality and the third inequality is by definition of the logarithm applied to both sides.  Let $\widetilde{x}_*$ denote the asymptotic limit for $\widetilde{x}_m$, namely $\widetilde{x}_m \rightarrow \widetilde{x}_*$ as $m\rightarrow \infty$. Then, taking the limit as $m\rightarrow \infty$, it follows from the above that
\begin{align}\label{eq:preUB}
\log\|\widetilde{x}_{*}-\widetilde{x}\|
&\leq \sum_{m=0}^\infty \log\|(I+\widetilde{F}_m)^{-1}\widetilde{F}_m \| + \log\|\widetilde{x}_0-\widetilde{x}\| 
\leq \sum_{m=0}^\infty \log \dfrac{\|\widetilde{F}_m\|}{1-\|\widetilde{F}_m\|}
+ C \nonumber \\
&=
\sum_{m=0}^\infty \left( \log \|\widetilde{F}_m\| - \log( 1-\|\widetilde{F}_m\|) \right) + C ,
\end{align}
where the second inequality is due to Equation~\eqref{eq:Fm-upperbound} and $C$ is a finite constant.

Taking the expectation of both sides of Equation~\eqref{eq:preUB} yields
\begin{align}\label{eq:UB}
\ex( \log\|\widetilde{x}_{*}-\widetilde{x}\| )
&\leq \ex\left( \sum_{m=0}^\infty \left( \log \|\widetilde{F}_m\| - \log( 1-\|\widetilde{F}_m\|) \right) + C \right) \nonumber \\
& = \sum_{m=0}^\infty \ex\left( \log \|\widetilde{F}_m\| - \log( 1-\|\widetilde{F}_m\|) \right) + C .
\end{align}
Consistent with the original analysis of Moler, we assume that the support for $\|\widetilde{F}_m\|$ is between $0$ and $1$, i.e., the distribution of $\widetilde{F}_m$ is such that $0 \leq \|\widetilde{F}_m\| < 1$.
Further assume that the distribution of $\|\widetilde{F}_m\|$ is symmetric around its mean with $\ex(\|\widetilde{F}_m\|)$ bounded away from $\frac{1}{2}$.
Then, due to the symmetry of $\log(x)-\log(1-x)$ for $0 < x < 1$, it follows that the second term in the expectation of Equation~\eqref{eq:UB} is greater than the first term, and therefore the expectation is negative and the summation tends to $-\infty$ since $C<\infty$.
To see this, note that $f(x) = \log(x)-\log(1-x)$ is monotonically increasing in $x$ and, for $\delta\in [0,\frac{1}{2})$, $f(\frac{1}{2}+\delta) = -f(\frac{1}{2}-\delta)$.
Let $p_m(x)$ be the probability distribution of the random variable $X_m = \|\widetilde{F}_m\|$ that is symmetric around its mean $\mu_m < \frac{1}{2}$. Then, $f(\mu_m-\delta) < -f(\mu_m+\delta)\leq 0$ and $\ex(f(X_m)) = \int p_m(x)f(x)dx \leq 0$ is in fact bounded away from $0$.  We therefore have established convergence in expectation of the iterative process under the assumptions above which implies that $\ex(x_*) = \widetilde{x}$.

Lastly, we establish a related result that exploits an important characteristic of analog computing devices~\cite{rasch2021flexible}, namely that the random variables $X_m = \|\widetilde{F}_m\|$ have different means $\mu_m = \ex(\|\widetilde{F}_m\|)$ whose average is asymptotically equal to zero; i.e., $\sum_{m=0}^{\ell-1} \mu_m/\ell \rightarrow 0$ in the limit as $\ell\rightarrow \infty$.
Consider the sequence of independent random variables $\{X_0,\ldots,X_{\ell-1}\}$ and assume that all of the random variables $X_m$ have the same variance, i.e., $\sigma_m^2 = \sigma^2$.
Define $s_\ell^2 := \sum_{m=0}^{\ell-1} \sigma_m^2 = \ell \sigma^2$ and thus $s_\ell = \sigma \sqrt{\ell}$.
Let us now consider
\begin{align}\label{eq:preLCLT}
\dfrac{1}{s_\ell} \sum_{m=0}^{\ell-1} (X_m-\mu_m)
&= \sum_{m=0}^{\ell-1} \left( \dfrac{X_m}{\sigma\sqrt{\ell}} - \dfrac{\mu_m}{\sigma\sqrt{\ell}} \right) 
= \sum_{m=0}^{\ell-1} \left( \dfrac{X_m}{\sigma\sqrt{\ell}} - \dfrac{\mu_m\sqrt{\ell}}{\ell \sigma} \right) \nonumber \\
&= \sum_{m=0}^{\ell-1} \dfrac{X_m}{\sigma\sqrt{\ell}} - \sum_{m=0}^{\ell-1} \dfrac{\mu_m}{\ell} \dfrac{\sqrt{\ell}}{\sigma} .
\end{align}
Assuming that $\sum_{m=0}^{\ell-1} \mu_m/\ell \rightarrow 0$ at a rate faster than $O(\sqrt{\ell})$ and taking the limit as $\ell\rightarrow\infty$ on both sides of Equation~\eqref{eq:preLCLT}, we obtain
\begin{equation}\label{eq:preLCLT2}
\lim_{\ell\rightarrow\infty} \dfrac{1}{s_\ell} \sum_{m=0}^{\ell-1} (X_m-\mu_m) = \lim_{\ell\rightarrow\infty} \sum_{m=0}^{\ell-1} \dfrac{X_m}{\sigma\sqrt{\ell}} .
\end{equation}
Further suppose that, for some $\delta > 0$, the following (Lyapunov) condition is satisfied:
\begin{equation*}
\lim_{\ell\rightarrow\infty} \dfrac{1}{s_\ell^{2+\delta}} \sum_{m=0}^{\ell-1} \ex[ |X_m - \mu_m|^{2+\delta} ] = 0 .
\end{equation*}
Then, from the Lyapunov central limit theorem~\cite{Billingsley} and Equation~\eqref{eq:preLCLT2}, we have that the sum of $X_m/(\sigma\sqrt{\ell})$ converges in distribution to a standard normal random variable as $\ell$ goes to infinity, namely
\begin{equation}\label{eq:LCLT}
\sum_{m=0}^{\ell-1} \dfrac{X_m}{\sigma\sqrt{\ell}} \quad \stackrel{d}{\rightarrow} \quad \mathcal{N}(0,1) , \qquad \mbox{  as } \ell\rightarrow\infty , 
\end{equation}
where $\mathcal{N}(0,1)$ denotes the standard normal distribution.

We have therefore established convergence in expectation of the iterative refinement process such that $\ex(x_*) = \widetilde{x}$, as long as the distribution of $\|\widetilde{F}_m\|$ is symmetric around its mean with $\|\widetilde{F}_m\|$ taking values in $[0,1)$ and $\ex(\|\widetilde{F}_m\|) < \frac{1}{2}$.  From Equation~\eqref{bnd:EF}, this implies the requirement that 
\begin{equation}\label{eq:parameterUB}
\ex(\|\widetilde{F}_m\|) \leq \kappa(\widetilde{A}) \left(n\sqrt{ \mu_K^2 + \sigma_K^2} \right) < \dfrac{1}{2}
\end{equation}
to ensure convergence in expectation of the iterative process.
Our asymptotic result in Equation~\eqref{eq:LCLT} suggests that we can drop $\mu_{K}$ as part of the overall iterative process, and thus this requirement can be further simplified over many iterations as
\begin{equation}\label{eq:parameterUB2}
\ex(\|\widetilde{F}_m\|) \leq \kappa(\widetilde{A}) \left(n\sqrt{\sigma_K^2} \right) = \kappa(\widetilde{A}) n\sigma_K < \dfrac{1}{2} .
\end{equation}
It is important to note, however, that Equations~\eqref{eq:parameterUB} and \eqref{eq:parameterUB2} only represent upper bounds on the parameters associated with $\|\widetilde{F}_m\|$, which can be (and are likely to be) quite loose.
Mathematically speaking, we really only need 
$\ex(\|\widetilde{F}_m\|) < \frac{1}{2}$ or $\|\widetilde{F}_m\| < \frac{1}{2}$ infinitely often.
In practice, we simply need to have $\|\widetilde{F}_m\| < \frac{1}{2}$ hold a sufficiently large number of times such that $\prod_{k=0}^m (I+\widetilde{F}_k)^{-1}\widetilde{F}_k < \frac{1}{2}$ for large $m$.

\subsection{Analysis of Preconditioned Richardson Iteration from an Analog Perrspective}
We next apply our error and convergence analysis above to the case of preconditioned Richardson iteration.
In particular,
analogous to the analysis in Appendix~\ref{app:IR-analog} for $\widetilde{K}_m$, we have that $\ex(x_{\ast})=x$ as long as $\ex(\|I-{M}A\|) < 1$.
With $E$ defined as in Section~\ref{sec-noise-analysis} to be
\[
E = M\odot(N^{Wm}+\mathbf{1}\otimes N^{Im} + N^{Om}\otimes\mathbf{1}) + N^{Wa}
\]
we then seek to establish bounds on $\|{E}\|$ such that
\begin{equation*}
    \|{E}\|<\dfrac{1-\|I-MA\|}{\|A\|}.
\end{equation*}
Analogous again to the analysis in Appendix~\ref{app:IR-analog} for $\widetilde{K}_m$, we assume ${E}$ to be a matrix whose elements are i.i.d.\ according to a general distribution $\cD_{E}(\mu_{E},\sigma_{E}^2)$ (not necessarily Gaussian) having finite mean $\mu_{E}$ and finite variance $\sigma_{E}^2$. 
Even though all matrix norms are equivalent in a topological sense, it is convenient and prudent for our purposes to choose matrix norms with which it is easy to work.
As in Appendix~\ref{app:IR-analog}, we shall continue to consider the $\max$ norm $\|\cdot \|_{\max}$, the Euclidean distance operator norm $\|\cdot \|_2$ and the Frobenius matrix norm $\|\cdot \|_F$.

Focusing on the norm 
$\|\cdot\|$ being either $\|\cdot\|_{\max}$ or $\|\cdot\|_2$, we follow our derivation above that leads to Equation~\eqref{eq:derivationK} and obtain
\begin{align}
\ex(\|{E}\|) \leq \ex(\|{E}\|_F)
& = n \sqrt{\mu_{E}^2 + \sigma_{E}^2} .
\label{eq:derivationE}
\end{align}
Similarly, from our analysis above leading to Equations~\eqref{eq:derivationK:var2} and \eqref{eq:derivationK:varub}, we respectively have
\begin{align}
\ex(\|{E}\|^2) \leq 
n^2 (\mu_{E}^2 + \sigma_{E}^2)
\label{eq:derivationE:var2}
\end{align}
and
\begin{equation}
\var(\|{E}\|)
\leq
n^2 \sigma_{E}^2 .
\label{eq:derivationE:varub}
\end{equation}
Our asymptotic result in Equation~\eqref{eq:LCLT} applied to $\|{E}\|$ suggests that we can drop $\mu_{E}$ as part of the overall preconditioned Richardson iteration process, and thus Equations~\eqref{eq:derivationE} and \eqref{eq:derivationE:var2} can be further simplified over many iterations as
\begin{equation}
\ex(\|{E}\|) \leq n \sigma_{E}
\qquad \mbox{ and } \qquad
\ex(\|{E}\|^2) \leq n^2 \sigma_{E}^2 .
\label{eq:E:1and2}
\end{equation}

Summing everything together, and assuming that write errors are the dominant 
source of noise, we can guarantee from Equation~\eqref{eq:derivationE} that the hybrid preconditioned Richardson iteration process will make progress in expectation during each iteration as long as 
\begin{equation*}
    \sqrt{\mu_{E}^2 + \sigma_{E}^2} < \dfrac{1-\|I-MA\|}{n\|A\|} ;
\end{equation*}
whereas Equation~\eqref{eq:E:1and2} suggests that such progress will be guaranteed in expectation over many iterations as long as 
\begin{equation*}
    \sigma_{E} < \dfrac{1-\|I-MA\|}{n\|A\|} .
\end{equation*}
Observe that, in both cases, the upper bound becomes larger when: ($a$) $\|I-MA\|\rightarrow 0$; ($b$) the spectral norm of the matrix $A$ becomes smaller; and/or ($c$) the size of matrix $A$ decreases.
The first component ($a$) is the only part that we can control by constructing a more accurate approximate inverse, while the second component ($b$) and third component ($c$) depend on the given matrix $A$.
Hence, depending upon the properties of the problem and the quality of the approximate inverse, we have from the above analysis an understanding of the suitability of the device properties.

At the same time, it is important to note that Equations~\eqref{eq:derivationE}~--~\eqref{eq:E:1and2} only represent upper bounds on the parameters associated with $\|{E}\|$, which can be (and are likely to be) quite loose.
Mathematically speaking, we really only need $\|{E}\| < \dfrac{1-\|I-MA\|}{\|A\|}$ infinitely often, and thus in practice we simply need to have $\|{E}\| < \dfrac{1-\|I-MA\|}{\|A\|}$ hold a sufficiently large number of times to realize convergence.

Similarly, for the bound on $E$ under the condition $M =A^{-1}+\Delta$, we have from the above analysis that a sufficient condition for convergence in expectation is given by
\[ \sigma_E < \frac{1}{n}(\|A\|^{-1}-\|\Delta\|) = \frac{1}{n}(\|A\|^{-1}-\|M-A^{-1}\|) . \]

\end{document}